\documentclass[iop]{emulateapj}

\usepackage{longtable}

\shorttitle{Magnetic reconnection during a flux emergence event}
\shortauthors{Guglielmino et al.}

\def\kms{$\textrm{km s}^{-1}$}

\begin{document}

\title{Multiwavelength Observations of Small-Scale Reconnection Events triggered by Magnetic Flux Emergence in the Solar Atmosphere}

\author{S. L. Guglielmino\altaffilmark{1}, L. R. Bellot
  Rubio\altaffilmark{2}, F. Zuccarello\altaffilmark{1}, G.
  Aulanier\altaffilmark{3}, S. Vargas Dom\'inguez\altaffilmark{4}
  and S. Kamio\altaffilmark{5} }

\altaffiltext{1}{Dipartimento di Fisica e Astronomia, Universit\`a di Catania, I 95123 Catania, Italy}
\altaffiltext{2}{Instituto de Astrof\'isica de Andaluc\'ia (CSIC), Apdo.\ de Correos 3004, E 18080 Granada, Spain}
\altaffiltext{3}{LESIA, Observatoire de Paris, CNRS, UPMC, Universit\'e  Paris Diderot, 5 place Jules Janssen, 92190 Meudon, France}
\altaffiltext{4}{Mullard Space Science Laboratory, RH5 6NT Holmbury St. Mary, Dorking, Surrey, UK}
\altaffiltext{5}{Max-Planck-Institut f\"ur Sonnensystemforschung, D 37191 Katlenburg-Lindau, Germany}

\email{salvo.guglielmino@oact.inaf.it}

\begin{abstract}

The interaction between emerging magnetic flux and the pre-existing ambient field has become a ``hot'' topic for both numerical simulations and high-resolution observations of the solar   atmosphere. The appearance of brightenings and surges during episodes of flux emergence is believed to be a signature of magnetic reconnection processes. We present an analysis of a small-scale flux emergence event in NOAA 10971, observed simultaneously with the Swedish 1-m Solar Telescope on La Palma and the \emph{Hinode} satellite during a joint campaign in September 2007. Extremely high-resolution G-band, H$\alpha$, and \ion{Ca}{2} H filtergrams, \ion{Fe}{1} and \ion{Na}{1} magnetograms, EUV raster scans, and X-ray images show that the emerging region was associated with chromospheric, transition region and coronal brightenings, as well as with chromospheric surges. We suggest that these features were caused by magnetic reconnection at low altitude in the atmosphere. To support this idea, we perform potential and linear force-free field extrapolations using the FROMAGE service. The extrapolations show that the emergence site is cospatial with a 3D null point, from which a spine originates. This magnetic configuration and the overall orientation of the field lines above the emerging flux region are compatible with the structures observed in the different atmospheric layers, and remain stable against variations of the force-free field parameter. Our analysis supports the predictions of recent 3D numerical simulations that energetic phenomena may result from the interaction between emerging flux and the pre-existing chromospheric and coronal field.

\end{abstract}


\section{Introduction}

Our understanding of the emergence of magnetic fields through the solar atmosphere has greatly improved over the last years thanks to new observations and numerical models \citep[see, e.g.,][and references therein]{Fan:09}. A large number of complex simulations, including radiative transfer effects and/or extending from the convection zone to the corona, have been carried out in order to shed light on the fate of the flux brought into the atmosphere by magnetic buoyancy, and on its possible interaction with the ambient field already present at the emergence site \citep[e.g.,][]{Galsgaard:07, Marti:08, Marti:09, Tortosa:09}.

The interaction between emerging flux regions (EFRs) and the ambient
field is usually believed to be responsible for perturbations in the
upper atmosphere. As a consequence of the reconnection between the
different flux systems, part of the magnetic energy stored in the
field lines is converted into heat and part into the kinetic energy of
impulsive phenomena, like jets and surges \citep{Shibata:95}.

Such a reconnection scenario has been investigated through MHD numerical simulations. They indeed show that the emerging magnetic flux may interact and reconnect with the pre-existing coronal field. The results obtained from the 2D models of \citet{Shibata:95,Shibata:96} have been confirmed by the more recent 3D models of \citet{Archontis:04,Archontis:05}, although the dynamics is significantly more complicated. The models suggest that flux emergence may be a relevant source of energy for the upper atmospheric layers and that magnetic reconnection between the EFR and the pre-existing field can lead to the ejection of chromospheric matter into the corona \citep{Archontis:05,Isobe:05,Marti:09}. At the reconnection site the temperature can reach $1-2 \times 10^{7} \;\textrm{K}$, and the plasma is emitted in the form of high-speed ($>100$ \kms) jets, at a height of $\sim 1.5-2 \;\textrm{Mm}$ above the base of the corona. The relative orientation between the two interacting flux systems seems to play an important role in the reconnection process, since it strongly influences the efficiency of the energy release and the consequent heating \citep{Galsgaard:07}. However, it ought to be realized that in all these ideal-MHD experiments reconnection occurs due to numerical resistivity, which is still several orders of magnitude larger than the physical resistivity of the solar plasma.

From an observational point of view, episodes of magnetic flux emergence are often accompanied by highly energetic phenomena in the chromosphere, like Ellerman bombs (EBs) and H$\alpha$ and H$\beta$ surges. EBs, discovered by \citet{Ellerman:17}, are small-scale brightenings observed in the wings of chromospheric lines, such as H$\alpha$ or \ion{Ca}{2} H. They have elliptical shapes and a typical size of $\sim 1\arcsec$  \citep{Georgoulis:02}. H$\alpha$ surges are straight or curved ejections of plasma with a filamentary structure. They may extend for up to $1-2 \times 10^{5} \;\textrm{km}$, reaching velocities of some tens of \kms{} \citep{KuroKawai:93}. EBs and surges have energies of the order of $10^{28} \;\textrm{erg}$ and lifetimes of $10 - 20$ minutes. They are thought to share a common origin in the ejection of chromospheric matter \citep[e.g.,][]{Matsumoto:08}.

In multiwavelength observations these features have been associated both in time and in space with coronal events found at flux emergence sites, such as UV/EUV brightenings \citep{Chae:99,Yoshimura:03} and X-ray jets \citep{Shibata:92}. Although several authors studied the relationship between surges and X-ray brightenings, the existence of a clear correlation is disputed. For instance, \citet{Canfield:96} observed an X-ray jet spatially separated from a surge, whereas \citet{Zhang:00} found an X-ray jet during the late stages of an H$\beta$ surge. On the other hand, it has been noticed that before the occurrence of a flare, coronal lines may show small-scale brightenings and non-thermal broadenings, which can be caused by magnetic reconnection, where flux is emerging and canceling \citep{Wallace:10}.

Recent observations have confirmed the results of MHD simulations in greater detail. \citet{Liu:04} presented a multiwavelength analysis of a chromospheric surge from the photosphere to the corona that supports the idea of magnetic reconnection in the low chromosphere. Also the high-resolution study of \citet{Brooks:07}, analyzing H$\alpha$ surges produced by the collision of small-scale magnetic flux concentrations of opposite polarity (moving magnetic features) around sunspots, showed that the chromosphere is heated up to coronal temperatures during magnetic field reconnection, in agreement with the findings of \citet{Shibata:96}. Furthermore, \citet{Moreno:08} carried out a numerical simulation of flux emergence which reproduced the three-dimensional structure of X-ray jets observed in coronal holes with the X-Ray Telescope \citep[XRT;][]{Golub:07} and the Extreme Ultraviolet Imaging Spectrometer \citep[EIS;][]{Culhane:07} aboard the \emph{Hinode} satellite \citep{Kosugi:07}.

Despite these findings, the relationship between the evolution of emerging magnetic fields and the fine structure of H$\alpha$ surges, \ion{Ca}{2} and EUV brightenings, and X-ray jets still has to be clarified. Small-scale changes in the topology of the emerging flux may lead to a dynamic energy release, so high-resolution observations are needed to shed some light on this issue.

Here we present a multi-wavelength analysis of extremely high spectral, temporal, and spatial resolution observations of a small-scale EFR, obtained during a coordinated campaign between the Swedish 1-m Solar Telescope \citep[SST;][]{Scharmer:03} and \emph{Hinode}. In a previous paper \citep{Guglielmino:08}, we used a subset of \ion{Ca}{2} H filtergrams, \ion{Na}{1} D1 magnetograms, and \ion{Fe}{1} $630.2 \;\textrm{nm}$ spectropolarimetric scans taken by Hinode's Solar Optical Telescope \citep[SOT;][]{Tsuneta:08} to study the rapid evolution of the EFR. Initially, the EFR showed upflows with horizontal magnetic fields in the central part and downflows with vertical field lines near the footpoints. As the magnetic flux increased, strong \ion{Ca}{2} H brightenings were detected at the contact line between the positive polarity of the EFR and a network element of opposite polarity, without clear counterparts in the G band. We interpreted these brightenings as the signature of chromospheric heating caused by the magnetic reconnection of the two flux systems.

Here we study the same EFR in the upper atmosphere, basing our analysis on the \ion{Ca}{2} H and H$\alpha$ observations acquired at the SST and the simultaneous measurements of XRT and EIS. These observations provide the most complete view of a flux emergence event ever obtained, as they cover all the atmospheric layers from the photosphere to the corona at the highest spatial resolution achievable with current telescopes.

We also perform a series of potential and linear force-free field (LFFF) extrapolations of the photospheric magnetograms to study the topology of the coronal field at the site of emergence and the interactions between the EFR and the ambient field during the emergence process.

In the next sections we describe the observations and the data reduction (Sect.~2), the evolution of the EFR at different heights (Sect.~3), and the topology of the magnetic field (Sect.~4). In Sect.~5 we suggest that the observed phenomena were caused by magnetic reconnection at low altitude in the solar atmosphere.

\section{Observations and data analysis}

The data analyzed in this paper were acquired during the \emph{Hinode} Operation Plan 14, a joint campaign between the solar telescopes of the Canary Islands and \emph{Hinode}. On 2007 September 30, the active region NOAA 10971 located at heliocentric coordinates ($178\arcsec$, $-61\arcsec$) was simultaneously observed with the SST and all the instruments aboard \emph{Hinode}.

The SST followed the active region from 8:45 to 10:05 UT, with a small gap between 9:31 and 9:36 UT. The observations were stopped when the seeing conditions became worse. \ion{Ca}{2} H ($\lambda = 396.85 \;\textrm{nm}$) and G-band ($\lambda = 430.56 \;\textrm{nm}$) images were recorded simultaneously, covering a field of view (FOV) of $63 \arcsec \times
64 \arcsec$ with a pixel size of $0.0338 \arcsec$. The full width at half maximum (FWHM) of the \ion{Ca}{2} H and G-band filters was $0.11\;\textrm{nm}$ and $1.2\;\textrm{nm}$, respectively. At the same time, the Solar Optical Universal Polarimeter \citep[SOUP;][]{Title:81} acquired filtergrams in the core of the H$\alpha$ line ($\lambda=656.29 \;\textrm{nm}$), together with high-resolution photospheric longitudinal magnetograms. The latter were derived from Stokes $I$ and $V$ measurements in the blue wing of \ion{Fe}{1} $630.25 \;\textrm{nm}$, at $\Delta \lambda = -5 \;\textrm{pm}$. The pixel size of the SOUP filtergrams is $0.0651 \arcsec$ and the FOV covers $63 \arcsec \times 60 \arcsec$. Additionally, broad-band images were recorded in the continuum of the \ion{Fe}{1} and H$\alpha$ lines to facilitate the alignment of the different datasets.

\begin{figure*}[!t]
\epsscale{1.05}
\plottwo{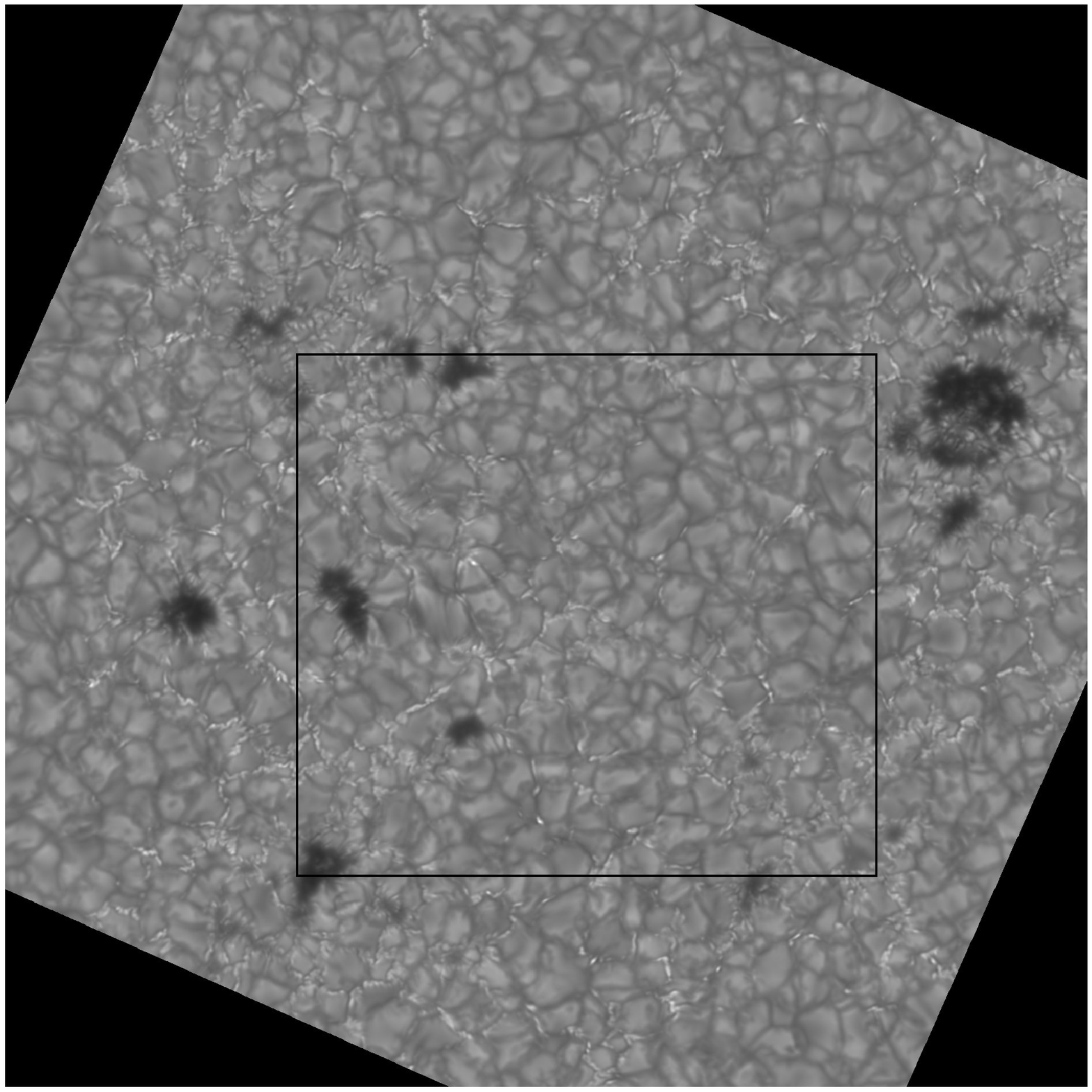}{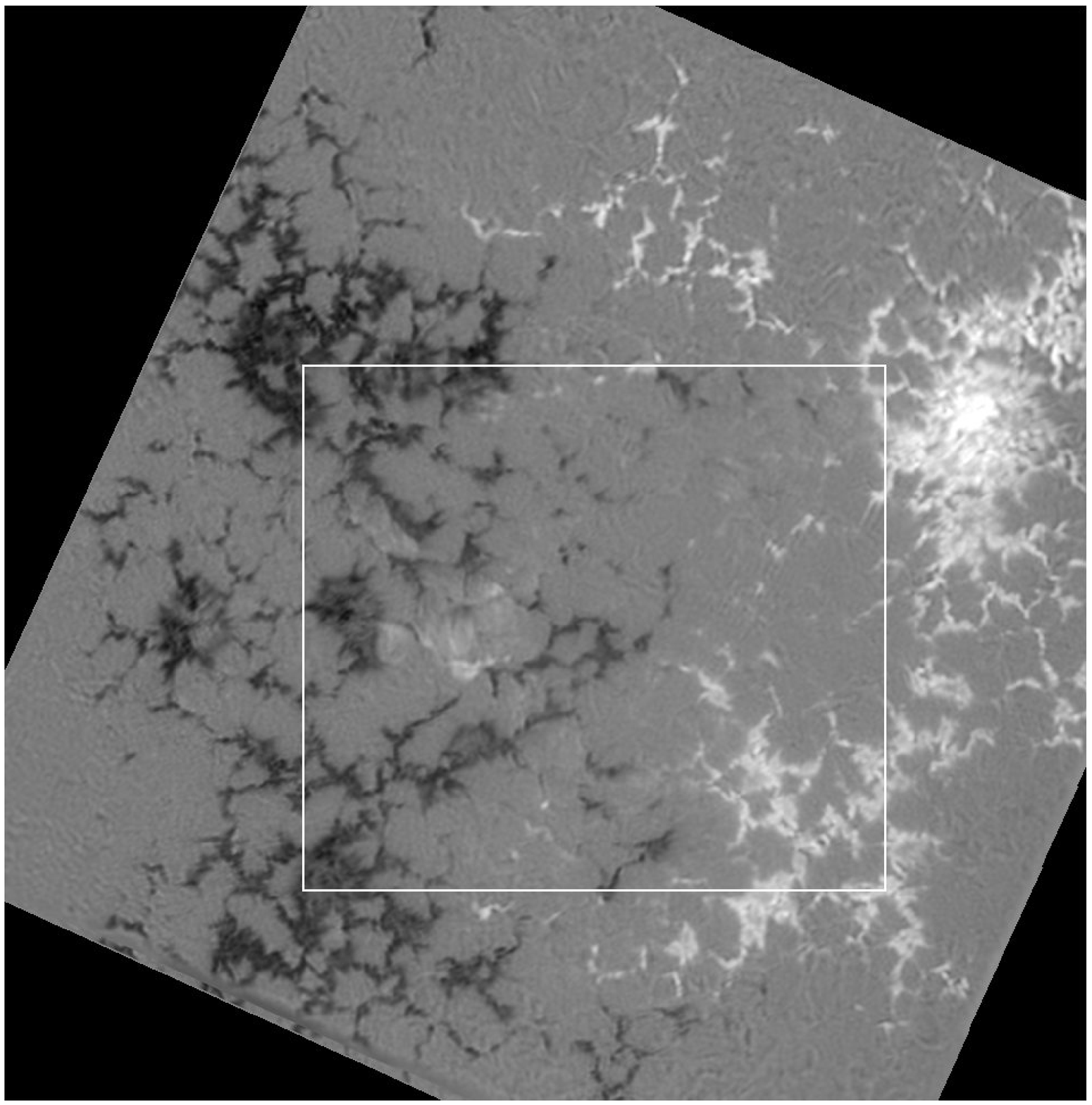}%
\caption{SST observations of NOAA 10971 on 2007 September 30 at 9:45 UT. {\em Left:} G-band filtergram, {\em Right:} \ion{Fe}{1} magnetogram. The squares indicate the part of the FOV shown in Fig.~\ref{fig3}. Throughout the paper, North is up and West to the right. \label{fig1}}
\end{figure*}

The observations were corrected for dark current and flat field before being processed with the Multi-Object Multi-Frame Blind Deconvolution technique \citep[MOMFBD;][]{vanNoort:05}. As a result, two time series of restored G-band and \ion{Ca}{2} H filtergrams with a cadence of 10 s were obtained, together with two time series of restored H$\alpha$ images and \ion{Fe}{1} magnetograms with a cadence of one minute. Further details about the MOMFDB processing can be found in
\citet{vannoortetal:08}. Thanks to the adaptive optic system of the SST and the MOMFBD restoration, the spatial resolution of $0\farcs2$ achieved at $630.2 \;\textrm{nm}$ is very close to the diffraction limit of the SST. Finally, the individual filtergrams were corrected for image rotation, co-aligned, and destretched. Figure~\ref{fig1} displays the entire FOV of a sample SST G-band image and a simultaneous \ion{Fe}{1} magnetogram showing NOAA 10971. 

The SST observations provide continuous coverage of the lower solar atmosphere. The G-band filtergrams trace the deep photosphere, while the \ion{Fe}{1} $630.25 \;\textrm{nm}$ measurements represent layers $150 - 300 \;\textrm{km}$ above $\tau=1$ \citep[e.g.,][]{2005A&A...439..687C}. The \ion{Ca}{2} H line forms at a temperature of $\sim 10000 \;\textrm{K}$ (low chromosphere), and the H$\alpha$ line at $\sim 15000 \;\textrm{K}$ \citep[mid-chromosphere; see, e.g.,][]{Vernazza:81}.

The method of analysis of the SOT data is discussed in
\citet{Guglielmino:08}. As regards other \emph{Hinode} observations,
we have used the \emph{SolarSoft} IDL package to obtain Level-1 data.
An area of $100 \arcsec \times 240 \arcsec$ was scanned by EIS with an
exposure time of $25$ sec per slit position, using the $2 \arcsec$
wide slit. The pixel size is then $1 \arcsec \times 2 \arcsec$, with
the larger size along the (EW) scan direction. Eight raster scans of
the active region were performed from 8:20 to 9:50 UT and from 10:20
to 11:50 UT, each one with a duration of $\sim 22$ minutes. The deep
exposures allow the detection of very weak signals, such as those
expected in the quiet Sun. The spectral ranges covered by the
observations contain a number of cool and hot emission lines (see
Table~\ref{tab:EIS}). The resolving power of EIS at these wavelengths
is $3000 - 4000$, as described by \citet{Young:07b}.

\begin{figure*}[!t]
\epsscale{1.15}
\plotone{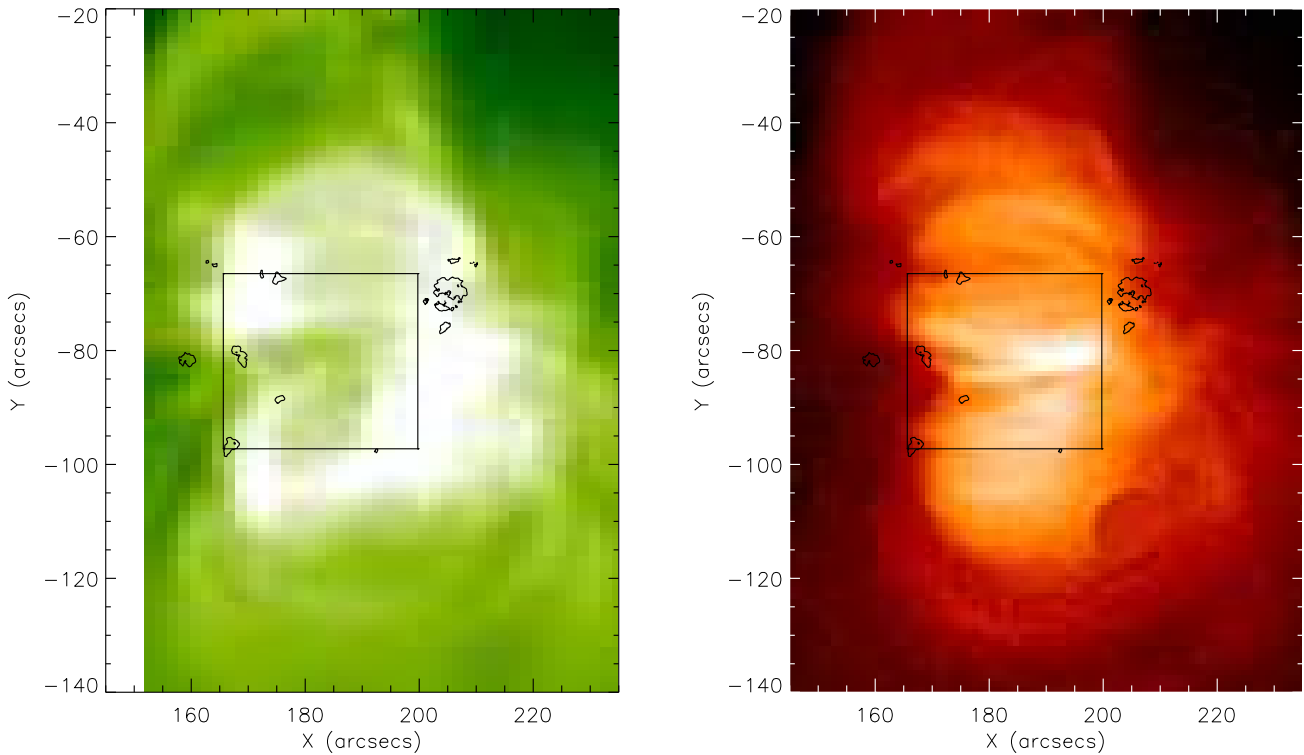}
\caption{\emph{Hinode} observations of NOAA 10971 at 9:45 UT. {\em Left:} map of radiance in \ion{Fe}{12} $195.12 \;\textrm{\AA}$ from the EIS raster scan started at 9:27 UT, {\em Right:} X-ray intensity through the C/polyimide filter. Both maps are shown in logarithmic scale, with the number of counts ranging from 0.05 to 3500 for EIS and from 10 to 2000 for XRT. The contours mark the position of the pores seen in the G band (Fig.~\ref{fig1}), and the squares indicate the part of the FOV displayed in Fig.~\ref{fig3}.\label{fig2}}
\end{figure*}

\begin{figure*}[!h]
\epsscale{1.01}
\plotone{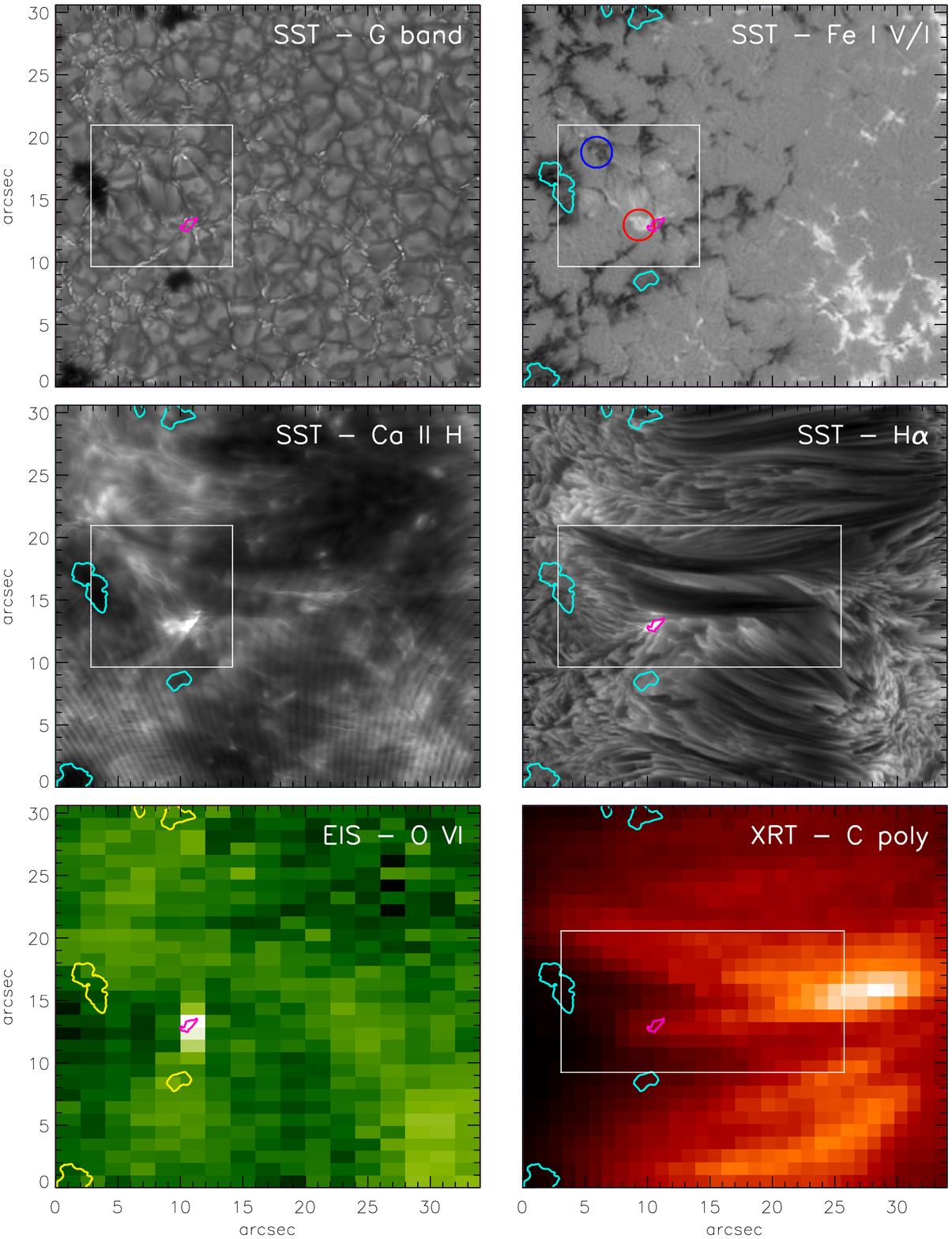}
\caption{EFR as observed at 9:45 UT, when it shows clear signatures in all the layers of the solar atmosphere. Displayed are a G-band filtergram (\emph{top left}), a \ion{Fe}{1} magnetogram (\emph{top right}), a \ion{Ca}{2} H line core filtergram (\emph{middle left}), and an H$\alpha$ line core filtergram (\emph{middle right}), all acquired at the SST. The bottom panels show the intensity of the \ion{O}{6} $184.12 \;\textrm{\AA}$ line from the EIS raster scan started at 9:27 UT (\emph{left}) and the X-ray intensity through the XRT C/polyimide filter (\emph{right}). The square, with a FOV of $\sim 12 \arcsec \times 12 \arcsec$, indicates the location of the EFR analyzed here. The emerging positive (negative) polarity is indicated with a red (blue) circle in the \ion{Fe}{1} magnetogram. For comparison, we overplot light blue (yellow on the EIS image) contours over a certain threshold of the G-band intensity to mark the location of the pores as seen in the photosphere, whereas purple contours indicate the site of the \ion{Ca}{2} H brightening. On the H$\alpha$ and X-ray images we overplot a rectangle with double size ($\sim 24 \arcsec \times 12 \arcsec$), to encompass the elongated chromospheric surge and the X-ray loop. The time sequences of G-band, \ion{Fe}{1}, \ion{Ca}{2} H and H$\alpha$ measurements are available as animations in the online journal
(AR10971\textunderscore{}20070930\textunderscore{}Gband.mov, AR10971\textunderscore{}20070930\textunderscore{}FeI6302.mov,
AR10971\textunderscore{}20070930\textunderscore{}CaII.mov, AR10971\textunderscore{}20070930\textunderscore{}Halpha.mov,
respectively). In the movies, major tick marks are separated by $10 \arcsec$. They show the same FOV, except the H$\alpha$ one which has a larger FOV in the horizontal direction to accommodate the surges. The contours represent the visible borders of the pores.\label{fig3}}
\end{figure*}

The EIS data were calibrated with the help of standard \emph{SolarSoft} procedures. In particular, we used the routine \texttt{eis\textunderscore{}prep}, which corrects for hot pixels and other bad data points and converts DN values into intensities in units of $\textrm{erg cm}^{-2} \textrm{s}^{-1} \textrm{sr}^{-1}\textrm{\AA}^{-1}$. The EUV spectra were corrected for the slit tilt and thermal drift using the \texttt{eis\textunderscore{}wave\textunderscore{}corr} routine. The long-wavelength detector was aligned with respect to the short-wavelength detector using a fixed displacement of $1$ pixel in the $x$-direction and 13 pixels in the $y$-direction, using the \ion{Fe}{12} $195.12 \;\textrm{\AA}$ radiance map as a reference. Moreover, we adopted the $\lambda$-dependent correction in the $y$-direction discussed by \citet{Young:09}, to whom we refer for further details about the EIS data calibration.

In order to extract line parameters from the EIS spectra, we implemented the MPFIT \emph{SolarSoft} procedures into the \texttt{fit\textunderscore{}line} routine, to automatically fit the calibrated spectra with Gaussians. In normal conditions, a Gaussian function is found to be a reasonable approximation to the line profiles \citep[e.g.,][]{Mariska:07}. The procedure allows the user to select the transition of interest and the corresponding wavelength range, and to deduce the integrated emission intensity (radiance), the background level (i.e., the continuum), the line centroid, and the FWHM. The Doppler velocity $v_{\rm D}$ can be easily calculated as
\begin{equation}
v_{\rm D} = c \frac{\lambda-\lambda_{0}}{\lambda_{0}},
\end{equation}
where $c$ represents the speed of light, $\lambda$ is the position of
the line centroid obtained from the fit, and $\lambda_{0}$ is the rest
wavelength of the transition. The values of $\lambda_0$ are
  extracted from the CHIANTI atomic database \citep{Landi:06,Dere:97},
  as listed by \citet{Brown:08}. The non-thermal velocity
  $v_{\mathrm{non-th}}$ was calculated by means of the
  \texttt{eis\textunderscore{}width2velocity} routine, which 
  removes the instrumental width $W$ and the known thermal velocity
  $v_{\mathrm{th}}$ of the line from the observed FWHM, using the relation
\begin{equation}
{\rm FWHM}^{2} = W^{2} + 4 \ln{2} \frac{\lambda^{2}}{c^{2}} (v_{\mathrm{th}}^2 + v_{\mathrm{non-th}}^{2})
\end{equation}
\citep{Harra:09}. The uncertainties in Doppler shift and
FWHM correspond to the formal errors of the Gaussian fit.

To derive the electron density we used a line ratio technique.
  The intensity ratio of density-sensitive line pairs is governed by
  the populations of the energy levels of the ion. The populations
  vary with the rate of ion-electron collisions, which depends on the
  electron density \citep[for more details, see][]{Phillips:08}. The
  relationship between the line ratio and the density was
  calculated using the CHIANTI database.

X-ray images were acquired through the Carbon polyimide (C/poly), the
Be thin, and the Be thick filters of \emph{Hinode}/XRT, with exposure
times of 1.03, 23 and 46~s, respectively. The cadence of the XRT
observations is 82~s.  These measurements cover a FOV of $384 \arcsec
\times 384 \arcsec$ with a pixel size of $1 \arcsec$. They have been
corrected for dark current, CCD bias, and cosmic rays using the
\texttt{xrt\textunderscore{}prep} routine.  The contamination spots
have been identified and removed by means of the
\texttt{xrt\textunderscore{}tup\textunderscore{}contam} procedure, and
finally the images have been corrected for pointing jitter.
Figure~\ref{fig2} shows NOAA 10971 as observed by EIS in the
coronal \ion{Fe}{12} 195.12 \AA\/ line and by XRT through the C/poly
filter.

\subsection{Co-alignment of the observations}

A critical step in multi-instrument observations is the precise co-alignment of the different datasets. Small offsets may also exist between images taken at different wavelengths with the same instrument. The internal offsets of SOT were evaluated by \citet{Shimizu:07} for the Broadband Filter Imager, while only pre-launch data are available for the \ion{Na}{1} line observed with the Narrowband Filter Imager.

\begin{figure*}[!t]
\epsscale{1.05}
\plotone{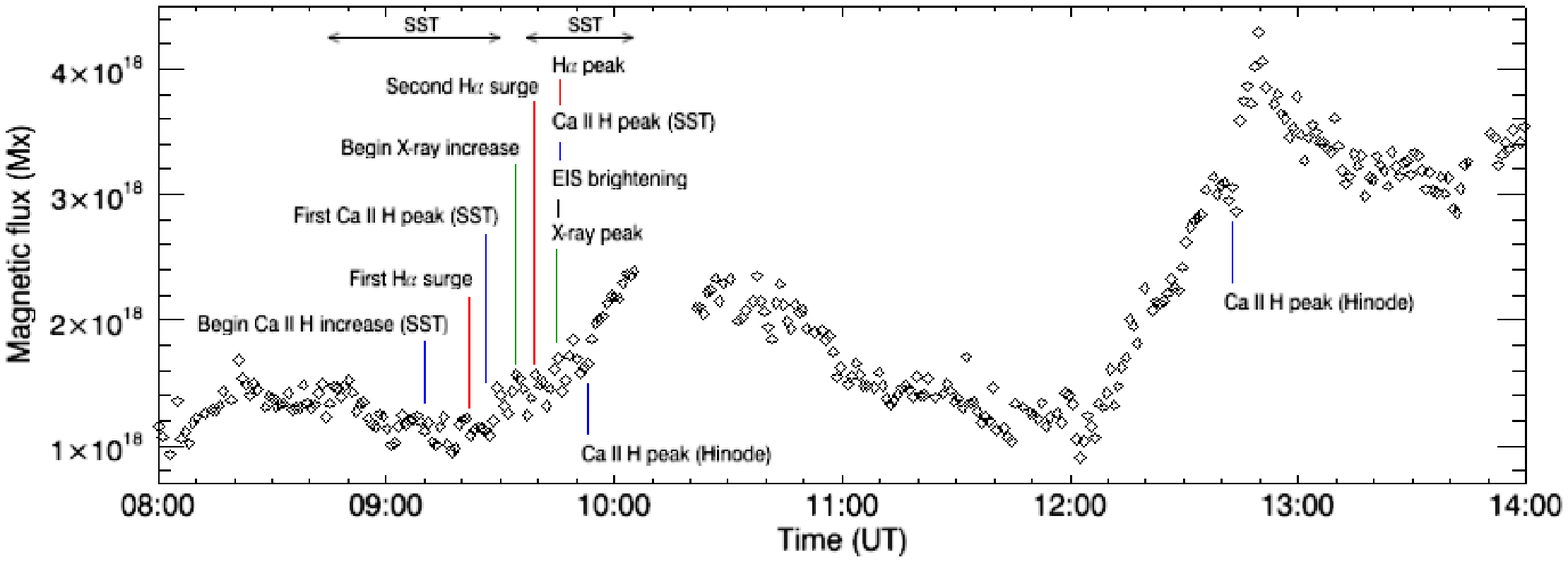}
\caption{Long-term evolution of the positive magnetic flux brought by the EFR, as computed from the \emph{Hinode} \ion{Na}{1} D1 magnetograms. The associated events are marked with labels. The horizontal segments indicate the periods with SST data.\label{fig4}}
\end{figure*}

The monochromatic images of each temporal series acquired at the SST
were aligned with respect to a reference time image using
cross-correlation techniques. To coalign the SOT and the SST images,
we used the G-band filtergrams from both instruments at the
reference time (9:30 UT), resampled to a pixel size of $0.055
\arcsec$. The displacement between them was obtained by cross
correlation, using the pores present in the FOV as fiducial points.
The same shift was applied to the simultaneous SST \ion{Ca}{2} H
images. To align the SST H$\alpha$ filtergrams and \ion{Fe}{1}
magnetograms, we cross-correlated the SOT G-band filtergrams with
the images taken in the continuum of the H$\alpha$ and \ion{Fe}{1}
lines. We estimate the accuracy of this procedure to be on the order
of $\pm 0\farcs1$.

	Unfortunately, the pointing information stored in the
  \emph{Hinode} FITS files is not sufficient to determine the 
  offsets between XRT, EIS, and SOT. Thus, we aligned the XRT
  filtergrams with respect to the SST H$\alpha$ channel. We compared
  the H$\alpha$ images with the XRT images, and not with the EIS
  scans, because the latter do not represent instantaneous snapshots
  and also because their spatial resolution is worse. The XRT image
  was superimposed on the corresponding H$\alpha$ filtergram by means
  of the \emph{iTools} IDL package. This application handles the image
  transparency and is very useful to check the correspondence of
  common structures, such as loops and bright regions. In particular,
  we aligned the X-ray coronal loops with the high-resolution
  H$\alpha$ features looking for a good overlap between them and for a
  position of the X-ray loops compatible with the endpoints observed
  in H$\alpha$. The alignment accuracy is comparable to the pixel size
  of XRT, i.e., $1\arcsec-2\arcsec$.

Finally, using cross-correlation and \emph{iTools} again, we
superimposed the EIS \ion{Fe}{12} $195.12\;\textrm{\AA}$ coronal
intensity map on the aligned XRT image, finding the location of the
EIS measurements with a precision of the order of $2\arcsec-3\arcsec$.

\section{Results}

An EFR appeared at about 8:00 UT at the internal edge of the main negative polarity of active region NOAA 10971
\citep[see][]{Guglielmino:08}. With a length of $\sim 8 \;\textrm{Mm}$ and a total photospheric flux of $1.4 \times 10^{19} \;\textrm{Mx}$, this EFR classifies as a small ephemeral region.

Figure~\ref{fig3} shows the EFR and its surroundings at 9:45 UT, from the photosphere (SST G-band filtergram) to the low corona (\emph{Hinode} XRT C/poly filtergram). It represents a global view of the emergence event in all the atmospheric layers covered by our observations.

The emergence of new flux into the photosphere is clearly seen in the magnetograms taken by \emph{Hinode}/SOT and the SST. Figure~\ref{fig4} displays the evolution of the positive magnetic flux in the upper photosphere as deduced from the \ion{Na}{1} D1 $589.6 \;\textrm{nm}$ measurements of \emph{Hinode} \citep{Guglielmino:08}.  There were two main episodes of flux emergence, beginning at around 9:20 UT and 12:00 UT. Only the former is covered by the SST observations.  Marked in the same figure are a number of events associated with the emergence process: brightenings in the chromospheric \ion{Ca}{2} H and H$\alpha$ lines, H$\alpha$ surges, and transient enhancements in cool EUV lines and X-rays. All these phenomena were observed during the first increase of the flux and span a broad range of heights in the solar atmosphere. Below we describe them in more detail.

\begin{figure*}[!t]
\epsscale{1.18}
\plotone{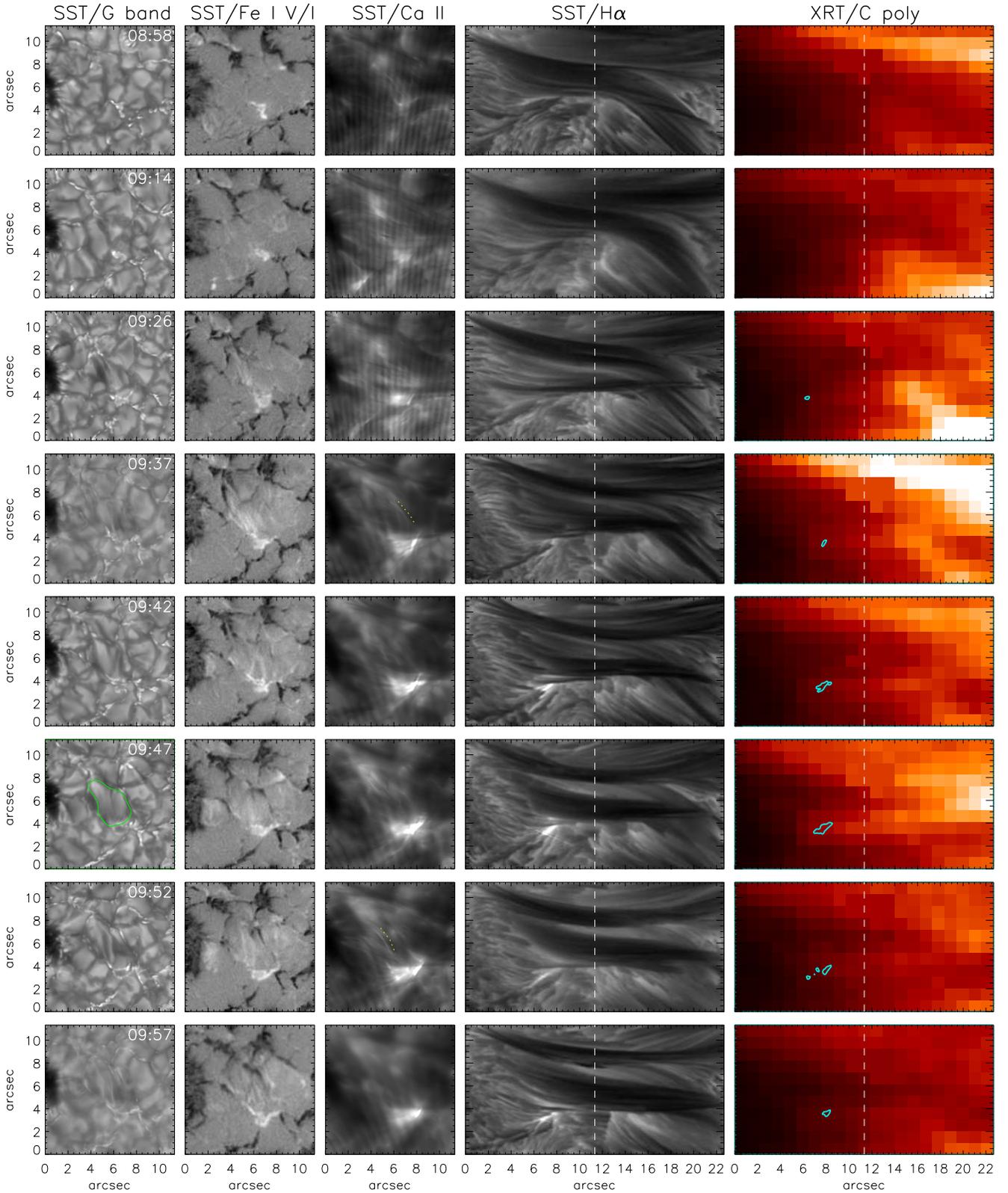}
\caption{Temporal evolution of the EFR between 8:58 and 9:57 UT. Displayed are simultaneous G-band filtegrams (\emph{first column}), \ion{Fe}{1} magnetograms (\emph{second column}), \ion{Ca}{2} H and H$\alpha$ filtergrams (\emph{third and fourth columns}), acquired at the SST, and XRT C/poly images (\emph{fifth column}), showing the EFR in the various atmospheric layers. The color tables are rescaled to the maximum intensity for each column. The FOV covers the squared/rectangular regions highlighted in Fig.~\ref{fig3}. The left half of the H$\alpha$ and X-ray images contains the FOV displayed in the other filtegrams. For comparison, blue contours in the X-ray filtergrams indicate the location of the \ion{Ca}{2} H brightenings.  The strong H$\alpha$ brightening and the enhanced emission of the X-ray loop cospatial to the \ion{Ca}{2} H brightening are clearly visible. At 9:47 UT the green contour in the G-band image indicates an anomalous granular cell at its maximum size. To help identify the fibrils that connect the two polarities of the EFR in the chromosphere, we have drawn yellow dotted lines parallel to them in some of the \ion{Ca}{2} H panels.\label{fig5}}
\end{figure*}

\subsection{Flux emergence in the photosphere}

The temporal sequence displayed in Fig.~\ref{fig5} summarizes the evolution of the EFR during the simultaneous observations of \emph{Hinode} and the SST. A comparison between the G-band images (\emph{first column}) and the magnetograms (\emph{second column}) initially shows bright points (BPs) at the position of the stronger flux patches of the EFR.  As new flux emerges, transient darkenings appear along the direction connecting the two main flux concentrations (see, e.g., the third and seventh rows). These dark lanes have a length of $3 \arcsec - 4 \arcsec$ and show BPs at their endpoints. Interestingly, a dark lane is visible in the continuum in the \emph{Hinode} spectropolarimetric scan taken at 09:18 UT, associated with upflows and horizontal fields \citep{Guglielmino:08}.

\begin{figure}[!t]
\epsscale{1.19}
\plotone{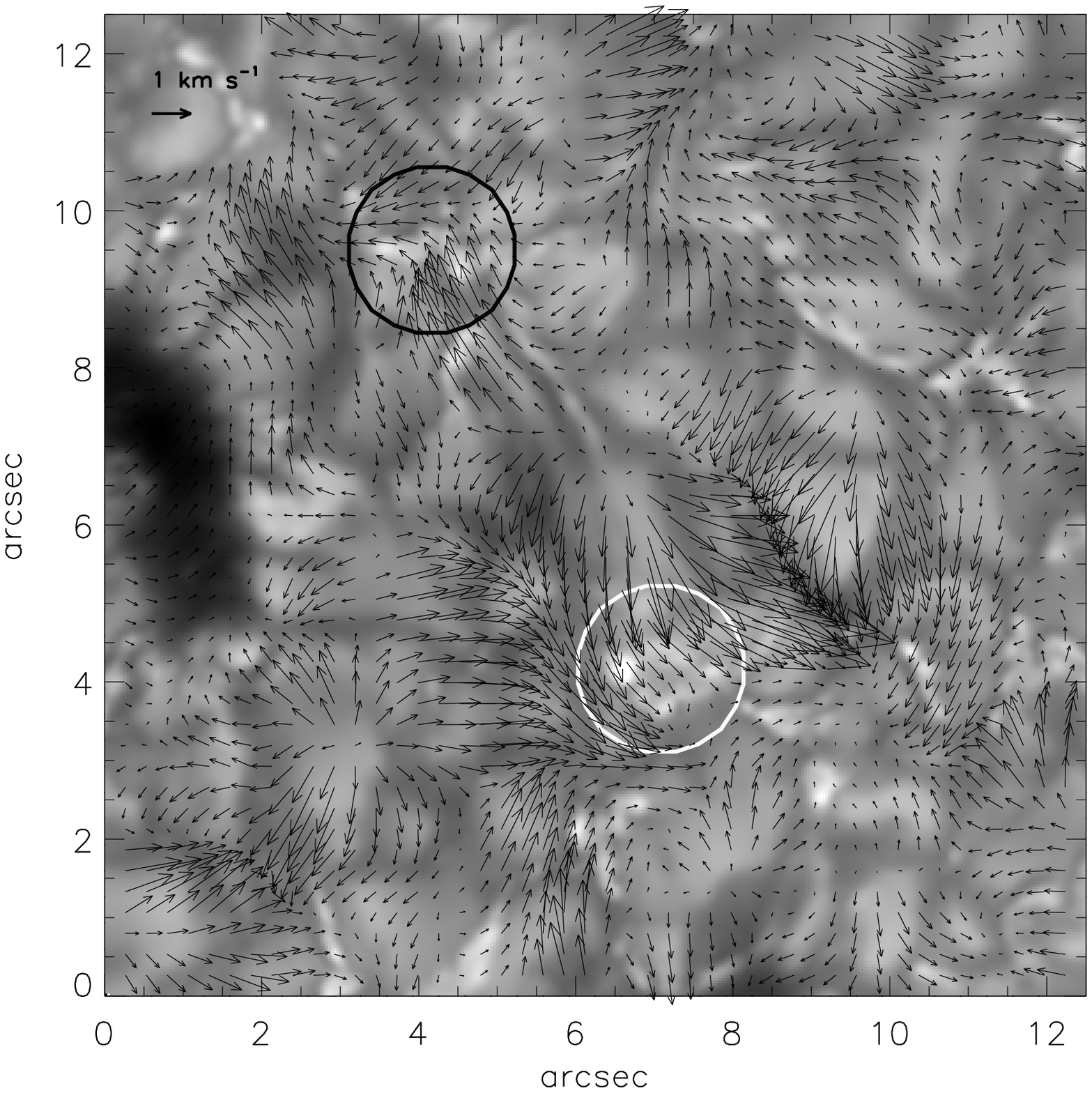}
\caption{Horizontal velocity field (15 minutes average from 9:15 to 9:30 UT), as derived by the LCT technique, overplotted on a SST G-band image acquired at 9:23 UT, with a FOV of $25 \arcsec \times 25 \arcsec$ centered on the EFR. The arrows indicate the direction and the magnitude of the horizontal velocity. The arrow in the left-top corner represents a velocity of $1$ \kms. The circles mark the position of the two main polarities of the EFR.\label{fig6}}
\end{figure}

The granulation pattern is also disturbed: the granular structures
between the two polarities of the EFR become larger and more elongated
than is typical in the quiet Sun, reaching a size of about $6
\;\textrm{Mm}^{2}$, almost twice as large as normal granular cells 
(see, e.g., the structure outlined by the green contour in the G-band
  image at 9:47 UT). The granules fragment after growing for $\sim
10$ minutes.

We have calculated the horizontal expansion velocity of the anomalous granules  using the Local Correlation Technique \citep[LCT; see][]{November:88}. The algorithm was applied to two subsets of SST G-band images, from 9:15 to 9:30 UT and from 9:37 to 9:52 UT, adopting a FWHM window of $5 \times 5$ pixels ($0\farcs17 \times 0\farcs17$) and a time interval of one minute. In Fig.~\ref{fig6} we display the horizontal velocity field obtained for the first subset, indicating the direction and magnitude of the motion with arrows. The average speed is about 3 \kms, with a peak of about 4 \kms{} near the BPs associated with the positive polarity of the EFR. We also recognize diverging motions from the center of the EFR toward the boundaries of the anomalous granules. These motions concentrate the magnetic field in the network patches observed around the EFR.

The dynamics revealed by our data confirms the findings of \citet{Strous:99}, now at smaller spatial scales. Similar dark lanes were detected previously by \cite{Otsuji:07} in \emph{Hinode} G-band and continuum images of an EFR that showed dark \ion{Ca}{2} H filaments. Transient darkenings and abnormal granulation have also been reported by \citet{Cheung:08}.

Our observations support the results of the numerical simulations of \citet{Cheung:07,Cheung:08} and \citet{Tortosa:09} concerning the photospheric evolution of EFRs. We indeed find upward velocities of $1.5$ \kms{} associated with strong horizontal fields of $\sim 1.5 \;\textrm{kG}$ \citep{Guglielmino:08}. However, there are some differences: for instance, the size reached by the anomalous granules is smaller than the simulated granules by at least a factor of 2.

We also note that the EFR is not formed by the emergence of a pair of coherent flux concentrations of opposite polarity. Rather, the flux appears as a series of small patches with mixed polarities, even if the EFR shows a net bipolarity (see the \ion{Fe}{1} movie in the online Journal). Such a pattern is extremely reminiscent of the sea-serpent shape of the emerging field lines seen in the simulations.

\subsection{Effects in the upper atmospheric layers}

Figure~\ref{fig5} shows that the EFR produced localized brightenings in the chromospheric \ion{Ca}{2} H and H$\alpha$ lines. They were first detected at 09:27 UT and 09:39 UT, respectively. The brightenings have an elliptical shape, with a width of $\sim 0\farcs15$ and a length of $\sim0\farcs7$, corresponding to about $100 \times 500 \;\textrm{km}^2$.

In the \ion{Ca}{2} H images (third column of Fig.~\ref{fig5}), the
connectivity between the two polarities of the EFR is revealed by
narrow bright filaments that join them and change continuously with
time (see frames 355-370 in the accompanying \ion{Ca}{2} movie).
The thickness of these filaments is about $2 - 3$ pixels,
corresponding to $\sim 0\farcs08$, right at the diffraction limit of
the telescope.  They are not to be confused with the interference
fringes aligned in nearly North-South direction, arising from
imperfect flatfield correction and amplified by the MOMFBD
process\footnote{We note that the only effect of such fringes is to
  increase the noise of the intensity measurements.}.  This
connectivity is not observed in the higher atmospheric layers,
probably because of the dominance of larger scale structures (see the
H$\alpha$ and the C/poly filtergrams in the fourth and fifth columns
of Fig.~\ref{fig4}).

The structure and the temporal evolution of the chromosphere is clearly visible in the \ion{Ca}{2} H movie. Plasma flows seem to be present in the region around the EFR, and the brightening just lies where some of these flows are converging. 

Three chromospheric surges, associated with the EFR and cospatial with some \ion{Ca}{2} H plasma flows, are recognized in the SST H$\alpha$ images starting at 9:23, 9:37, and 9:52 UT. The surges appear southward of an arch filament system (AFS) connecting the opposite polarities of the active region. The loops of the AFS show continuous morphological variations and their footpoints are anchored in the magnetic network outlining the boundaries of supergranular cells (see Fig.~\ref{fig1}).  The lack of spectroscopic information does not allow us to estimate the downflow/upflow motions and the plasma speed in the loops, which is usually reported to be on the order of $10-15$ \kms{} \citep{Zuccarello:09}.

\begin{figure*}[!p]
\epsscale{1.18}
\plotone{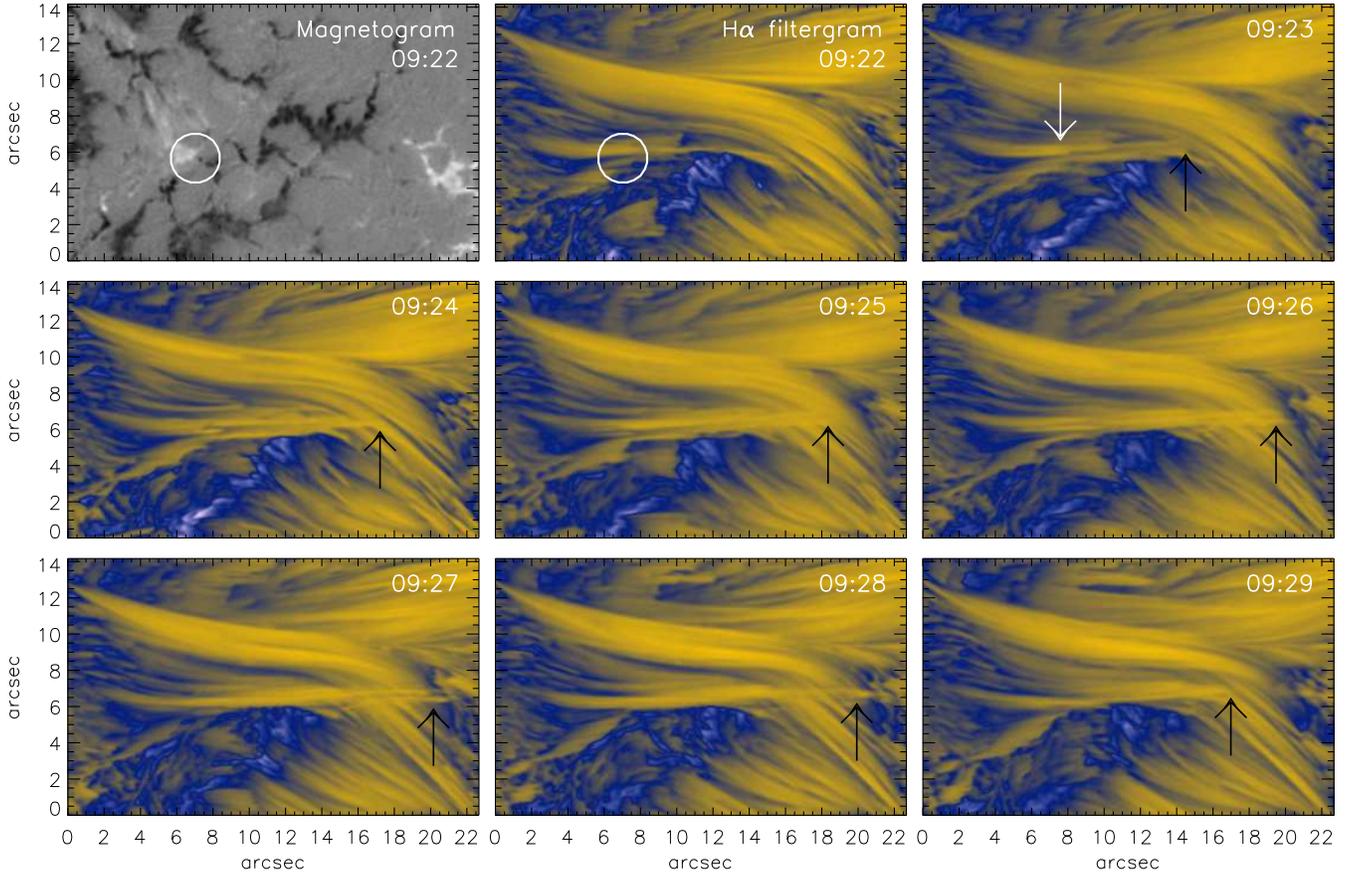}
\caption{Time sequence of H$\alpha$ line core filtergrams, showing the development of the first surge associated with the EFR. The images have a cadence of one minute. For comparison, the first panel shows the \ion{Fe}{1} magnetogram corresponding to the first H$\alpha$ image (9:22 UT). The sharp boundary of the chromospheric surge (black arrows) allows us to easily follow the temporal evolution of the structure. The white arrow indicates the \texttt{Y}-shaped structure seen since the initial stages of surge. A full animation including this and the other surges can be found in the online Journal.\label{fig7}}
\end{figure*}

The temporal sequence of H$\alpha$ images displayed in Fig.~\ref{fig7} shows the evolution of the first chromospheric surge with enhanced colors\footnote{The second and third surges are smaller and more difficult to identify, hence we will not consider them further. However, we note that the third surge appears to be associated with a strong H$\alpha$ brightening that peaked at 9:50 UT, some $2\arcsec$ north of the main brightening of Fig.~\ref{fig3}.}. To facilitate comparisons, the initial photospheric magnetogram is given in the first panel. Thanks to the extremely high spatial resolution of the measurements, the filamentary structure of the surge is clearly visible. A \texttt{Y}-shaped feature at the base of the surge is recognized at 9:23 UT (indicated by the white arrow). At 9:40 UT, its cross-point location coincides with the site of the H$\alpha$ and \ion{Ca}{2} H brightenings (compare the fourth and fifth rows of Fig.~\ref{fig5}).

\begin{figure*}[!p]
\epsscale{1.175}
\plotone{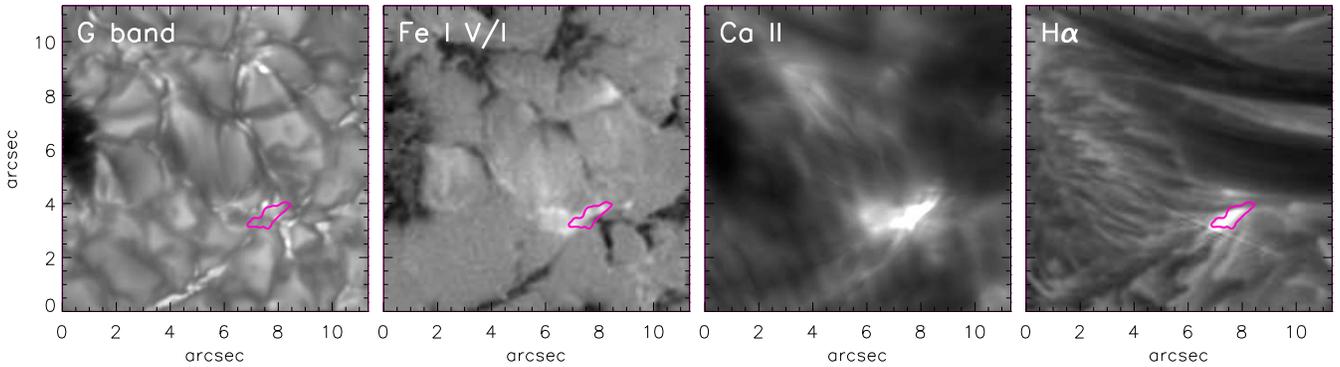}
\caption{EFR as observed in the photosphere/chromosphere at 9:47 UT. The FOV is the same as in Fig.~\ref{fig5}. Purple contours indicate the site of maximum \ion{Ca}{2} H intensity. As can be seen, the \ion{Ca}{2} H and H$\alpha$ brightenings coincide with BPs and with the contact region of the approaching polarities.\label{fig8}}
\end{figure*}

The sharp edge of the surge allows us to estimate the projection of its velocity in the horizontal direction. To that end we have considered the distance between the centroid of the EFR and the west edge of the surge: there is an initial impulsive phase during which the distance increases (9:23 -- 9:28 UT), followed by a decay phase (9:29 -- 9:30 UT). We have performed a parabolic fit to the impulsive phase: the (projected) speed after the first minute of the ejection is 53 \kms, whereas the average speed during the impulsive phase is 34 \kms, in agreement with previous estimates \citep[e.g.][]{Brooks:07}. The uncertainty of these values is about $\pm 2$ \kms.

Figure~\ref{fig8} is a close-up of the 9:47 UT observations displayed in Fig.~\ref{fig5}. It shows that the \ion{Ca}{2} H and H$\alpha$ brightenings occurred in the contact region between the emerging positive polarity and a pre-existing network element of negative polarity. Contours indicating the maximum \ion{Ca}{2} H line-core intensities are overplotted for reference. BPs can be seen at the position of the small-scale flux patches, but there are no indications of any heating process around the contact line in the photospheric G-band filtergrams.

\begin{figure}[!t]
\epsscale{1.15}%1.175
\plotone{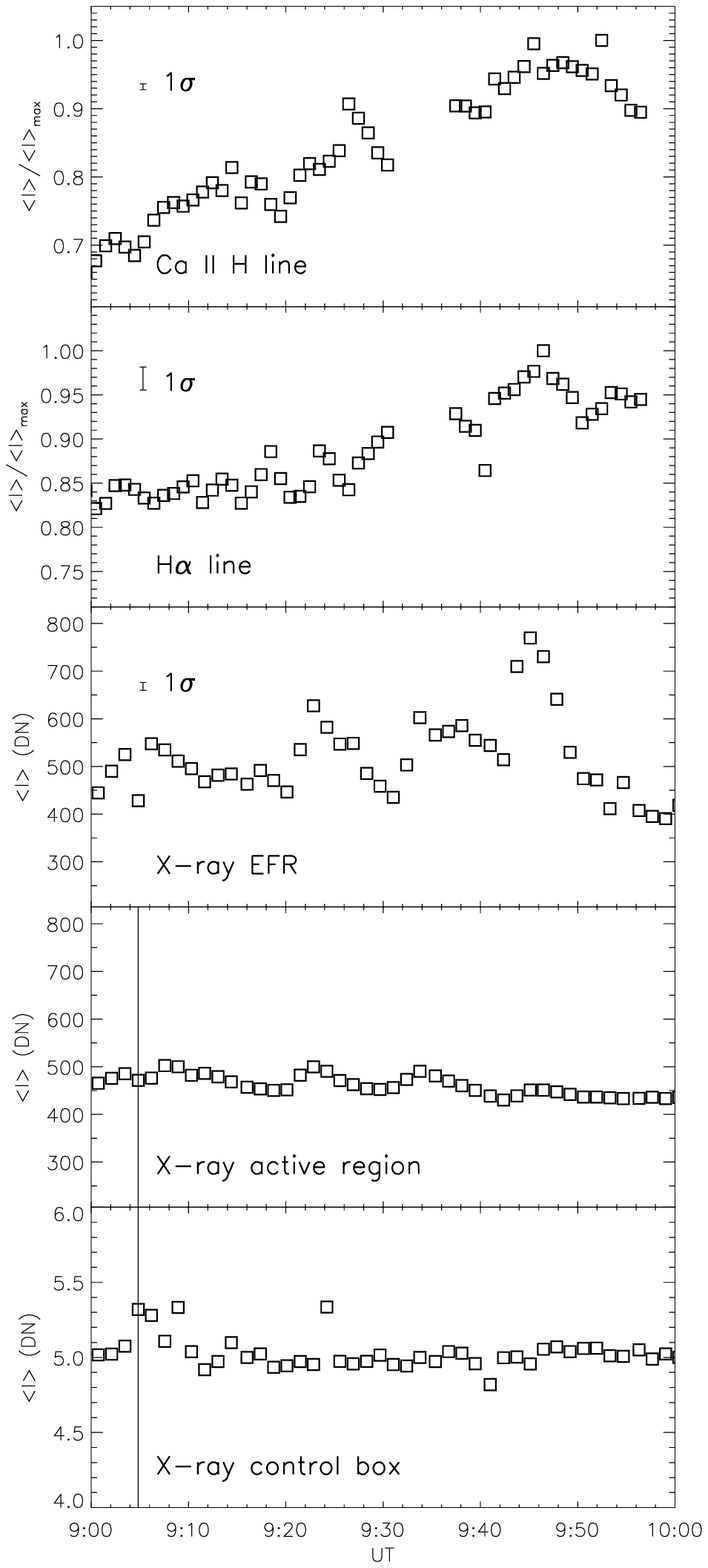}
\caption{Temporal evolution of the \ion{Ca}{2} H and H$\alpha$ intensities, normalized to their maximum value, and the X-ray intensities (EFR, active region,  and control box). There was a simultaneous peak in the \ion{Ca}{2}, H$\alpha$, and X-rays curves around 9:45 UT. The vertical lines in the X-ray plots indicate the time when \emph{Hinode} left the South Atlantic Anomaly. The 1$\sigma$ marks indicate the amplitude of the error bars.\label{fig9}}
\end{figure}

Figure~\ref{fig5} shows that the site of the \ion{Ca}{2} H brightening coincides with the footpoint of an X-ray loop with enhanced emission, which could be a jet. Coronal variations during the emergence event were observed only in the images taken with the C/poly filter. This filter is sensitive to temperatures down to $\log T = 6$, while the thin and thick Be filters react to much hotter temperatures \citep{Golub:07}. Thus, it appears that the energy released during the emergence process was not sufficient to heat the plasma well above $\sim 10^6 \;\textrm{K}$.

\begin{figure}[!t]
\epsscale{1.15}%1.175
\plotone{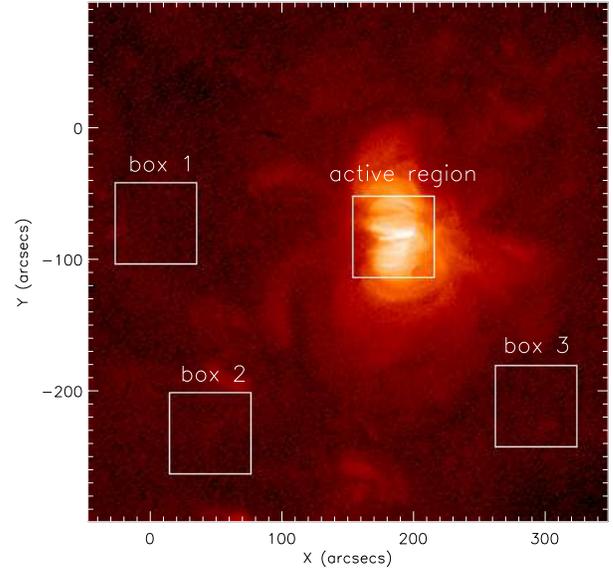}
\caption{Full FOV of the XRT measurements with the C/polyimide filter
  at 9:45 UT. The maps is shown in logarithmic scale between 1 and
  2000 counts. The box containing the active region and the three
  control boxes used for the plots in Fig.~\ref{fig9} are shown. All
  the boxes have a size of $63\arcsec \times 63\arcsec$.\label{fig10}}
\end{figure}

We plot in Fig.~\ref{fig9} the temporal evolution of the maximum
\ion{Ca}{2} H, H$\alpha$, and X-ray brightness at the emergence
site\footnote{To obtain these values, in each frame we have
  averaged the four pixels of maximum intensity within a $5\arcsec
  \times 5\arcsec$ box centered on the positive polarity patch
    of the EFR. The box is the same for all the data sets.}. The
\ion{Ca}{2} H intensity and later the H$\alpha$ intensity show a
continuous increase beginning at 9:05 and 9:27 UT, respectively,
whereas the X-ray enhancement is more abrupt. Despite this difference,
the maximum X-ray intensity was observed at the time of the strongest
\ion{Ca}{2} H and H$\alpha$ brightenings (around 9:45 UT). In
  Fig.~\ref{fig9} we also display the mean X-ray intensities within a
  box containing the entire active region (see Fig.~\ref{fig10}). It
  looks quite flat, confirming that the X-ray increase in the EFR was
  not associated with a general increase of the background emission in
  the active region. The bottom panel of Fig.~\ref{fig9} shows the
mean XRT intensity in the three control boxes of
  Fig.~\ref{fig10}. This curve demonstrates that the X-ray
enhancements in the EFR were solar in origin and not due to variations
of the XRT transmission or cosmic ray events, except perhaps for the
small peaks seen during the crossing of the South Atlantic Anomaly and
a single peak at around 9:25 UT. However, the latter represents a
variation of less than 10\% in the control box, while in the EFR it
corresponds to a variation of 50\%. The X-ray peak at 9:45 UT is not
associated with any increase in the control box.

\subsection{EIS spectroscopic analysis}

Observations taken with EIS fill the gap between the H$\alpha$ and the XRT observations. EIS observed a brightening in the transition region and in the lower corona above the H$\alpha$ brightening at 9:45 UT, during the raster scan started at 9:27 UT. 

\begin{figure*}[p]
\includegraphics[angle=90,scale=.6]{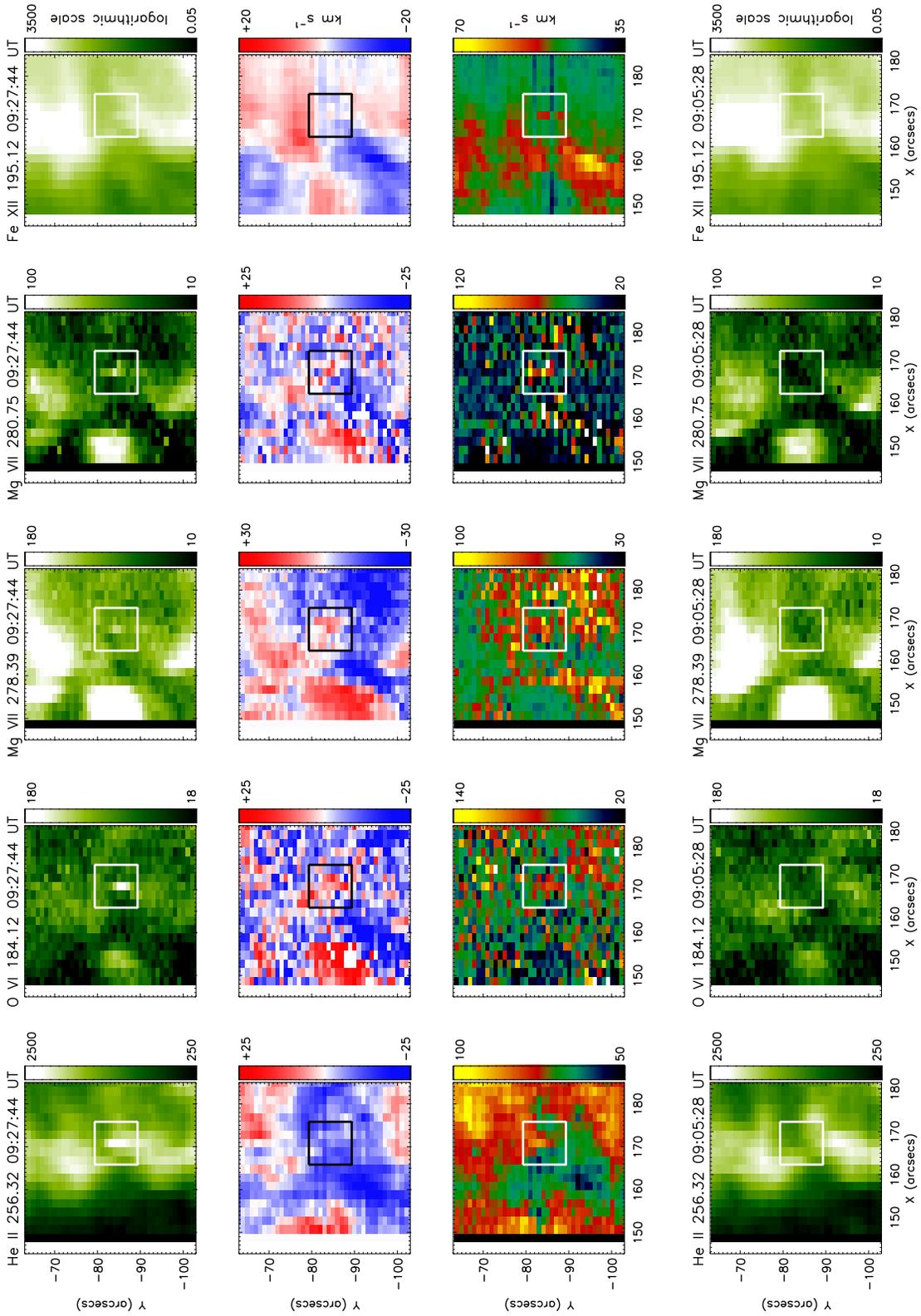}
\caption{Maps of radiance (\emph{first row}), Doppler velocity (\emph{second row}), and line width (\emph{third row}) obtained for five EUV lines observed by EIS during the scan started at 9:27 UT. The squares indicate the region where the brightenings have been observed. For comparison, we also report the radiance map  (\emph{fourth row}) for the same five lines during the previous scan, started at 9:05 UT.\label{fig11}}
\end{figure*}

Figure~\ref{fig11} displays, from top to bottom, radiance, Doppler velocity, and FWHM maps in \ion{He}{2} 256.32~\AA\/, \ion{O}{6} 184.12~\AA\/, \ion{Mg}{7} 278.39~\AA\/, \ion{Mg}{7} 280.75~\AA\/, and \ion{Fe}{12} 195.12~\AA\/. For comparison, the bottom row shows the radiance maps obtained during the previous scan, started at 9:05 UT, where the brightenings are totally absent.

Since the brightening was observed at only one slit position $2\arcsec$-wide, an adjacent slit position to the west is regarded as unperturbed medium in the following study. EUV enhancements were detected in the full temperature range $10^{4.7} - 10^{6.1}$ K. One of the most pronounced brightenings occurred in the \ion{Mg}{7} 278.39~\AA\/ line: its radiance was $2.5$ times brighter than the adjacent background. The variation of the Doppler shift is smaller than the uncertainty of the measurements ($6$ \kms), and thus no significant velocity change was detected. The width of the \ion{Mg}{7} 278.39~\AA\/ line increased from $0.09 \pm 0.01$ \AA{} in the background to $0.15 \pm 0.02 $ \AA{} at the position of the brightening. Assuming an ion temperature of $10^{5.8} \;\textrm{K}$ for this line and the EIS instrumental width of $0.06$ \AA, the non-thermal velocity in the brightening is estimated to be 81 \kms, compared to 34 \kms{} for the background. The observed increase of the FWHM without a net Doppler shift implies unresolved upflows and downflows within the EIS pixel.

\begin{figure*}[!t]
\epsscale{1.18}
\plotone{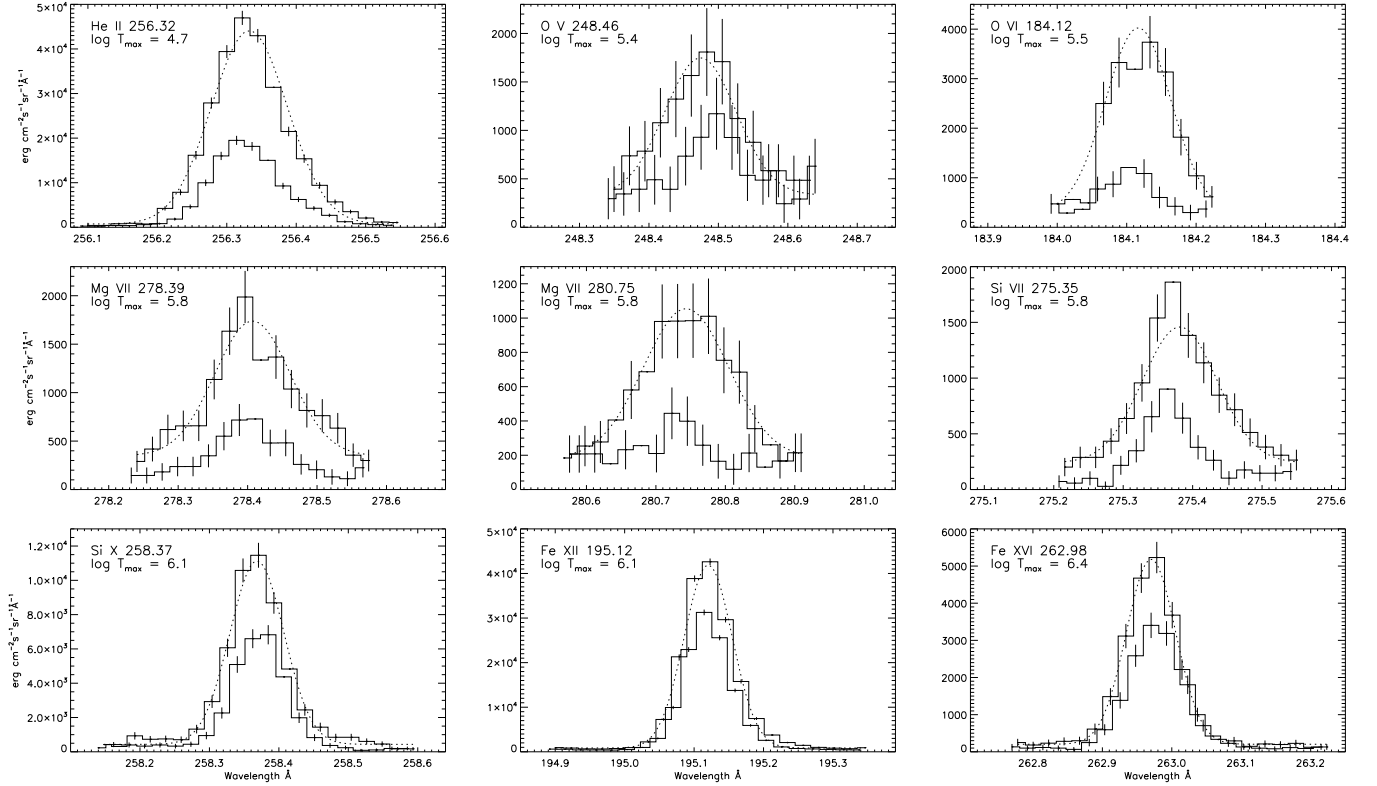}
\caption{Intensity profiles of nine EUV lines with increasing ion temperatures, observed by EIS before (\emph{solid line}) and during the brightenings (\emph{solid thick line}). Dotted lines represent Gaussian fits to the profiles at 9:45 UT.\label{fig12}}
\vspace*{1em}
\end{figure*}

In Fig.~\ref{fig12} we plot the observed profiles at the site of the
brightening for a sample of EUV lines, before and during the intensity
enhancement. For strong emission lines such as \ion{Si}{10}
258.37~\AA\/, a Gaussian function reproduces the observed profile
within the measurement errors. Weak emission lines exhibit deviations
from a Gaussian shape, mainly because of the low signal-to-noise ratio
of the spectrum. However, the small enhancements seen in the wings of
\ion{Mg}{7} 278.39~\AA\/ could also indicate high speed flows at that
temperature.

The electron density is derived from the density-sensitive line pair
\ion{Mg}{7} 278.39~\AA\/ and \ion{Mg}{7} 280.75~\AA, in accordance
with \citet{Young:07a}. Even if ionization equilibrium were not
  reached in the transient brightening, the density diagnostics would
  not be affected as we use only the ratio of lines from the same ion,
  which is determined by collisional excitation processes. Since the
  line ratio technique is based on the total intensity, i.e., the
  integral of the emission profile, the exact line shape does not
  affect the density analysis either \citep{Phillips:08}. The estimated electron density at the
position of the brightening is $10^{9.6\pm 2} \;\textrm{cm}^{-3}$,
which is roughly $2$ times as large as the background value of
$10^{9.2\pm 2} \;\textrm{cm}^{-3}$.  The density increase may be the
signature of chromospheric evaporation. It should be noted that the
inferred physical quantities are averages over the EIS pixel, hence
the real variations could have been larger. The H$\alpha$ filtergrams
indeed show that the width of the chromospheric surge was smaller than
$1\arcsec$.

\begin{deluxetable*}{lccccccc}[!b]
\tablecaption{EUV parameters \label{tab:EIS}}
\tablehead{
\colhead{Ion } & \colhead{ $\lambda_0$ (\AA)} & \colhead{ $T_{\rm max}$ (K)} & $I_{\rm b}$ & \colhead{ $v_{\rm D}$ (\kms)} & \colhead{ FWHM (\AA)} & \colhead{ $v_{\mathrm{non-th}}$ (\kms)} &  \colhead{ $v_{\mathrm{th}}$ (\kms)} 
}
\startdata
\ion{He}{2} 	& 256.32 	& $10^{4.7}$ & 2.9 & $-7  	\pm 1\phn $  &  $ 0.135    \pm  0.002$  &	$\phn85 \pm \phn2$  & 14.4 \\
\ion{O}{5} 	  &	248.46	& $10^{5.4}$ & 5.1 & $\phn5 \pm 10 $     &  $ 0.13\phn \pm   0.04$  &	$\phn87 \pm 30$  		& 15.6 \\
\ion{O}{6} 	  &	184.12	& $10^{5.5}$ & 4.9 & $\phn5 \pm 8\phn$ 	 &  $ 0.11\phn \pm   0.02$  &	$\phn86 \pm 30$		  & 17.7 \\
\ion{Mg}{7} 	&	278.39	& $10^{5.8}$ & 3.6 & $12  	\pm 5\phn  $ &  $ 0.12\phn \pm   0.02$  &	$\phn75 \pm 10$ 		& 20.8 \\
\ion{Mg}{7} 	&	280.75	& $10^{5.8}$ & 9.7 & $16  	\pm 10 $ 		 &  $ 0.18\phn \pm   0.05$  &	$110   	\pm 30$ 		& 20.8 \\
\ion{Si}{7} 	&	275.35	& $10^{5.8}$ & 3.4 & $20  	\pm 5\phn  $ &  $ 0.15\phn \pm   0.02$  &	$\phn90 \pm 10$ 		& 18.5 \\
\ion{Si}{10} 	&	258.37	& $10^{6.1}$ & 1.6 & $-3  	\pm 2\phn  $ &  $ 0.084		 \pm  0.004$  &	$\phn34 \pm \phn5$	& 28.6 \\
\ion{Fe}{12}	&	195.12	& $10^{6.1}$ & 1.4 & $1.0		\pm 0.6$ 		 &  $ 0.083    \pm  0.001$  &	$\phn52 \pm \phn1$ 	& 20.0 \\
\ion{Fe}{16}	&	262.98	& $10^{6.4}$ & 1.4 & $-8 		\pm 2\phn  $ &  $ 0.083    \pm  0.004$  &	$\phn32 \pm \phn5$	& 28.0
\enddata
\tablecomments{\footnotesize{Rest wavelength $\lambda_0$ from
      the CHIANTI database, ion temperature $T_{\rm max}$,
    ratio between the integrated intensity during and before
      the brightening $I_{\rm b}$, Doppler velocity $v_{\rm D}$, line
    width at half maximum FWHM, non-thermal velocity
    $v_{\mathrm{non-th}}$, and (tabulated) thermal velocity
    $v_{\mathrm{th}}$ in the pixel where the brightening
      occurs, for a sample of EUV lines observed by EIS}.}
\end{deluxetable*}

The physical quantities extracted from the other EUV lines are
reported in Table~\ref{tab:EIS}. The \ion{He}{2} 256.32~\AA\/
  line is blended with three coronal lines in the red wing, but we
  cannot correct for that\footnote{For instance, we do not have
    observations in the spectral range 258.69 -- 262.65~\AA, where the
    \ion{Si}{10} 261.04~\AA\/ line provides a fixed intensity ratio
    with the \ion{Si}{10} 256.37~\AA\/ line, one of the lines blended
    with \ion{He}{2} 256.32~\AA\/.}. Nevertheless, the results should
  be reliable, because in disk observations  \ion{He}{2}
  256.32~\AA\/ contributes more than 80\% to the blend
  \citep{Young:07b}.

  The \ion{Fe}{12} 195.12~\AA\/ line is one of the
    hottest in the sample and shows only a faint
  brightening.  Its parameters are consistent with quiet conditions in
  the corona and do not seem to be affected by the EFR. However, the
  continuum intensity of the line is strongly enhanced at that
  position, by a factor of 2.  This could indicate the presence of
  plasma with higher temperature than usual.

  The other lines show large non-thermal widths and downflows in the
  brightening, except for the lowest \ion{He}{2} 256.32~\AA\/ line and
  \ion{Fe}{16} 262.98~\AA\/ which show upward motions. Given the
  uncertainty of the velocity measurements, the downflows cannot be
  regarded as significant. Only the upflows detected in \ion{He}{2}
  256.32~\AA\/ and \ion{Fe}{16} 262.98~\AA\/ are well above the
  velocity error, but in these cases the brightening is embedded in a
  region of strong upflows. Thus, it is unclear whether the emergence
  event produced velocity changes or not.

  As can be seen in Fig.~\ref{fig12}, the brigthness enhancement
  decreases with the ion temperature, i.e., with increasing height.
  This confirms the conclusion drawn from the absence of X-ray signals
  in the hottest XRT filters, namely that the plasma did not reach
  very high temperatures.

\section{Analysis of the magnetic configuration}

In order to obtain information on the coronal connectivity above the
EFR, we performed a series of potential and linear force-free field
extrapolations. The coronal loops observed in the EUV and
  X-ray images of Fig.~\ref{fig2} reveal a non-stressed magnetic
  configuration: no sigmoid structure is evident and the loops in the
  arcade appear to be parallel to each other. Such a configuration is
  indicative of a low-energy state that should be well represented by
  these extrapolations.

We used the BLFFF code from the FRench Online MAGnetic
Extrapolations (FROMAGE) service, which is based on the FFT method of
\citet{Alex:81}. The magnetic field \textbf{B} was computed for $z>0$
using $B_{z}=B_{\parallel} \cdot \cos\psi$ as photospheric boundary
conditions at $z=z_0$, where $B_{\parallel}$ is the observed $V/I$
value, binned for computer memory reasons, and $\psi$ is the angle
between the \emph{z} axis and the radial direction at the solar disk
center, taking into account projection effects.

The extrapolations were performed for the SST magnetograms obtained at
8:58 UT and 9:30 UT, and the \emph{Hinode}/SOT magnetograms taken at 8:10,
8:42 and 8:57 UT. The geometry and topology of the extrapolated
magnetic fields are not sensitive to the exact value of the conversion
factor between $V/I$ and the magnetic field, provided this factor
is nearly constant throughout the observed FOV.

In the extrapolations, the deprojected magnetogram was inserted into a
wider region of $L^2 \;\textrm{Mm}^2$ (using various $L$ between 200
and 340 Mm) with zero vertical component $B_z$, so as to minimize the
aliasing effects resulting from the periodic boundary conditions along
the \emph{x} and \emph{y} axes. The total $B_z$ at $z=z_0$ was
developed into Fourier eigenmodes, using a resolution of $1024 \times
1024$. Hence, the original resolution of the magnetogram was degraded
by a factor $\sim 5.1$. This degradation does not change the overall
topology and field line connections. Using the linear force-free field
equation
\begin{equation}
{\bf \nabla}\times\mathbf{B} = \alpha\mathbf{B},
\end{equation}
the $B_{x,y}$ components at $z=z_0$, as well as all three components
of \textbf{B} at various $z>z_0$, were calculated according to the
value of the force-free parameter $\alpha$. In our study, $\alpha$ was
a free parameter with (maximum and minimum) resonant values
$\alpha^\prime=\pm2\pi/2L$. The potential fields were calculated using
$\alpha=0$. The output vector \textbf{B} was then written in a
$301 \times 301$ mesh, with the smallest cell being $\textrm{D}=0.17$
Mm around the EFR, and slowly increasing toward the edges of the box
in the \emph{x} and \emph{y} directions. For $z>0$, \textbf{B} was
calculated in 201 layers above the plane of the magnetogram, the
smallest interval in \emph{z} being 0.13 Mm, and slowly increasing up
to $z=z_0+100$ Mm.

\subsection{Fan-spine topology}

\begin{figure*}[!t]
\epsscale{0.98}%1.175
\vspace{.5cm}
\plottwo{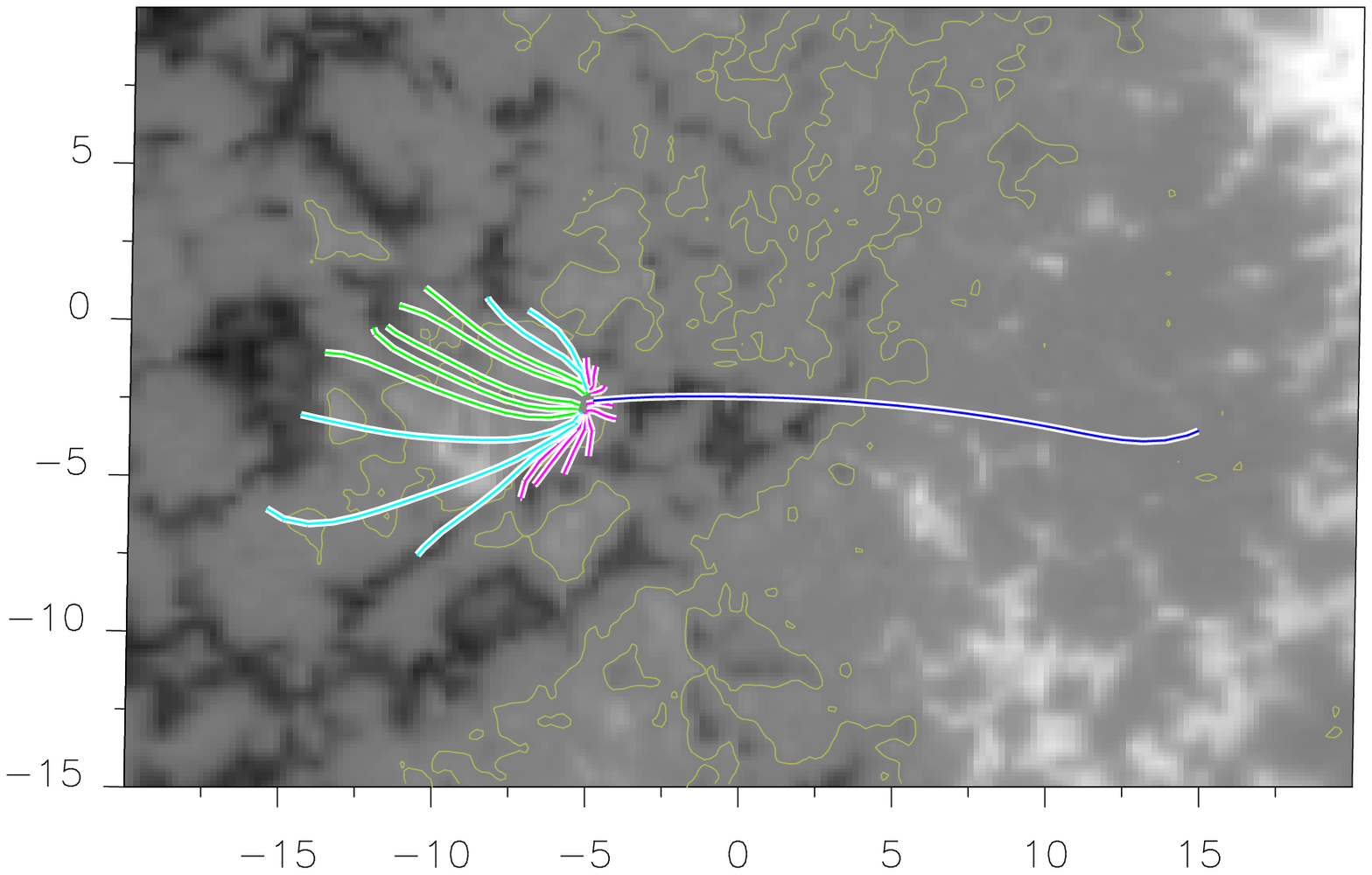}{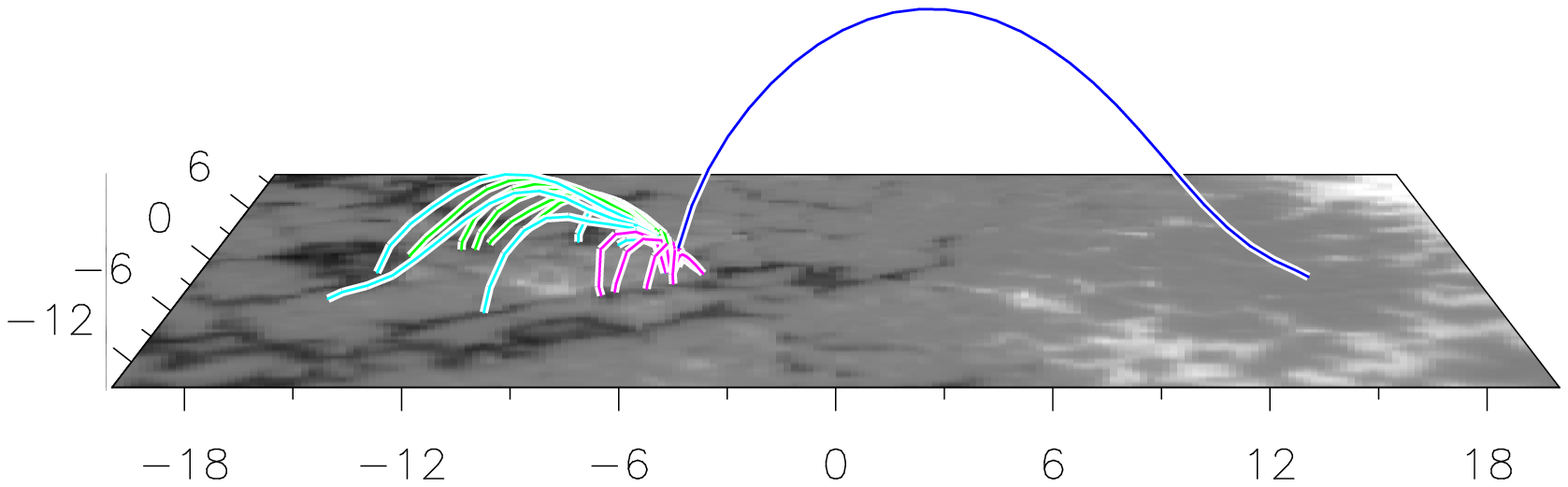}\hspace{.5cm}(a)\vspace{.8cm}%
\plottwo{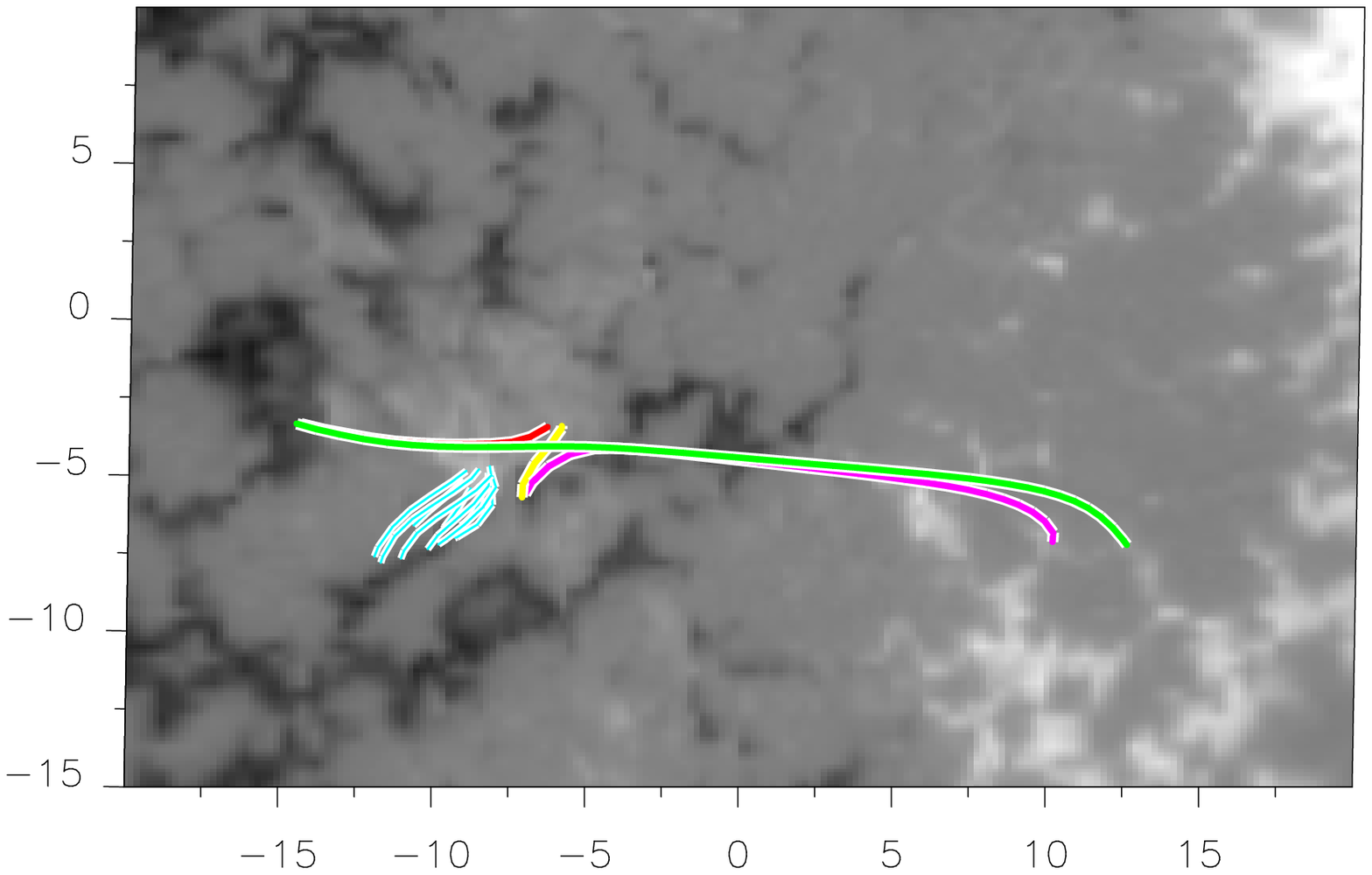}{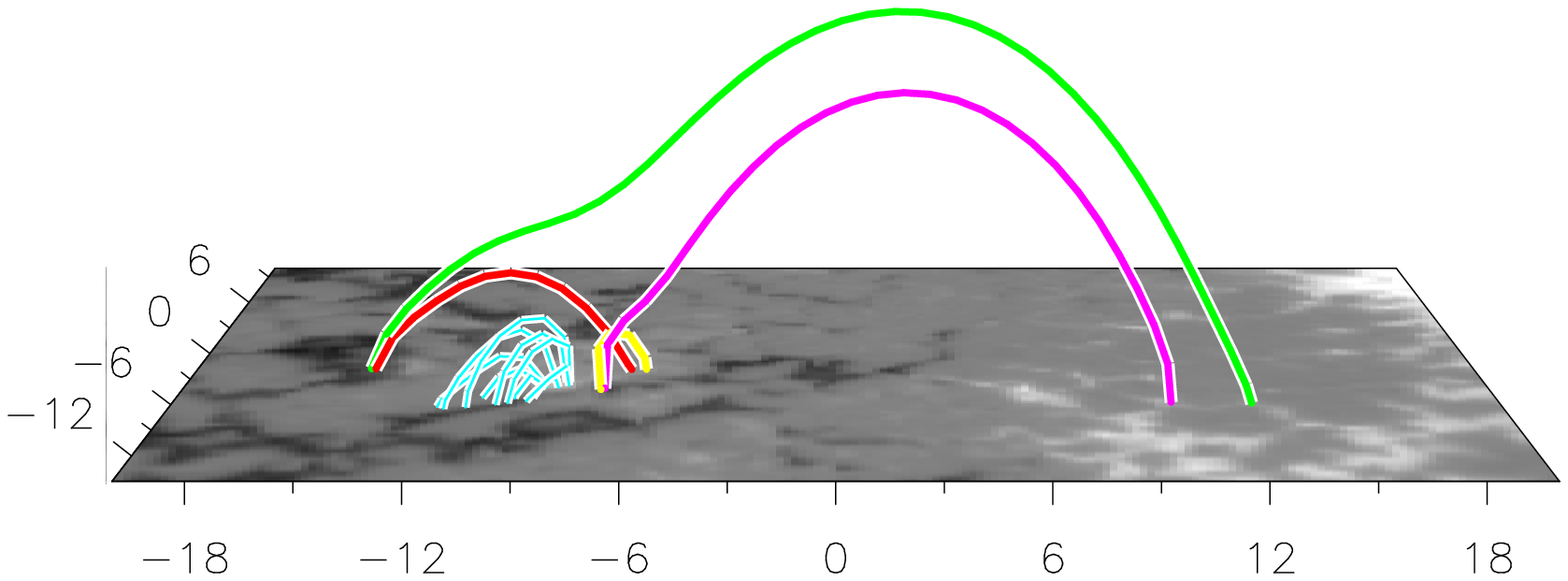}\hspace{.5cm}(b)\vspace{.2cm}
\caption{Magnetic field configuration of the active region around the
  EFR. The left column shows projections as viewed from Earth. The
  right column shows lateral perspectives looking toward solar north,
  with the \emph{z} axis stretched by a factor two for better
  visibility of magnetic field lines. \emph{Top:} Separatrix field
  lines emanating from the null point in the potential field
  extrapolation of the SST magnetogram acquired at 9:30 UT. The dark
  blue line is the spine field line, and other colored lines are fan
  field lines. The background image represents $B_z$ at $z=z_0$, and
  yellow contours indicate polarity inversion lines. \emph{Bottom:} Four
  selected field lines crossing each other slightly south from the
  null point, with neighboring NW-SE low altitude field lines (cyan).
  The cyan/red/yellow field lines are enclosed by the fan surface.
  \label{fig13}}
\end{figure*}

\begin{figure*}[!t]
\epsscale{0.8}%1.175
\plotone{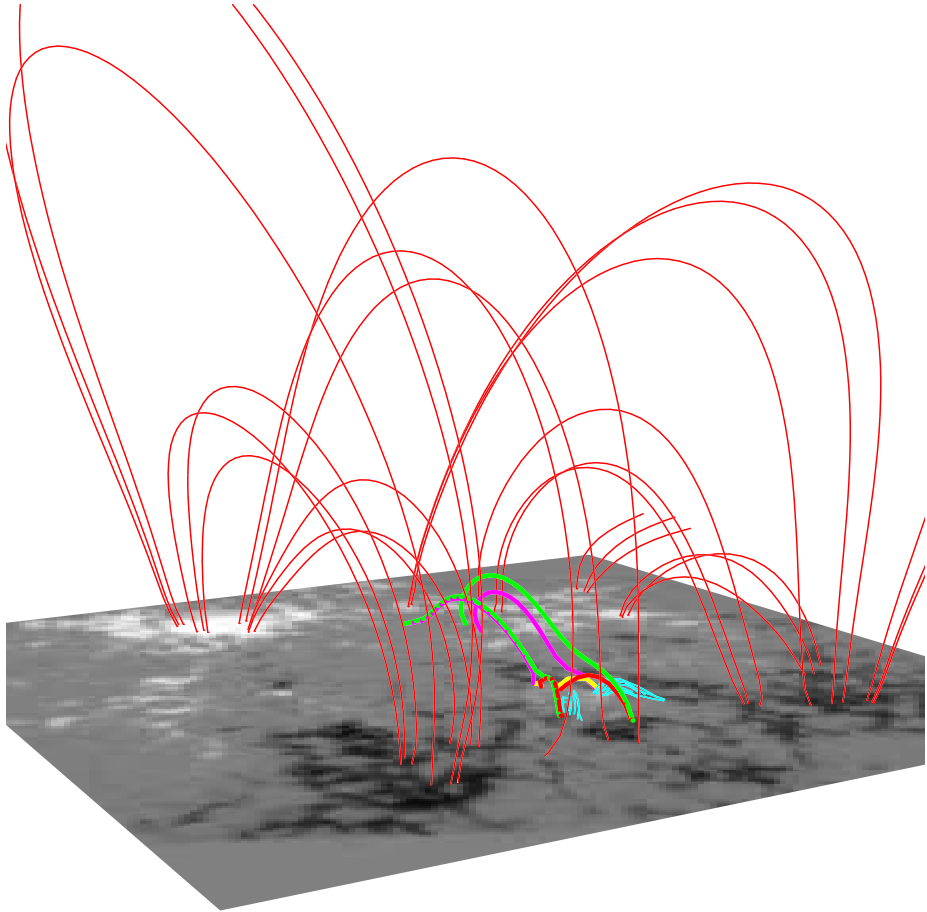}
\caption{Large-scale view of the potential extrapolation shown in Fig.~\ref{fig13}. Thin red lines represent the field lines connecting the main polarities of the active region. Thick colored lines show the same groups of field lines displayed in Fig.~\ref{fig13}b. The path of the long thick red and green lines closely follows that of the spine of the null point.\label{fig14}}
\end{figure*}

We performed extrapolations of the five magnetograms listed above, for
several values $|\alpha|\le3/4\, \alpha^\prime$. While the orientation
of the field lines connecting the main polarities of the active region
is sensitive to $\alpha$-variations, the global topology of the
magnetic field as well as the overall orientation of the field lines
above the EFR change very little from one extrapolation to the next.
So, for simplicity, we hereafter focus on the results of the potential
field extrapolation of the SST magnetogram taken at 9:30 UT (shown in
Figs.~\ref{fig13} and~\ref{fig14}).

An asymmetric null point is observed above the EFR. The strength of
the large-scale magnetic field of the active region makes this null to
be located on the west side of the main positive polarity of the EFR,
slightly NW from the location of the \ion{Ca}{2} H and H$\alpha$
brightenings.  Depending on the extrapolation, the altitude of the 3D
null point is typically $150-400 \;\textrm{km}$ above that of the
magnetogram.  Since the formation height of the \ion{Fe}{1} $630.25
\;\textrm{nm}$ line is $z_0\simeq250 \;\textrm{km}$ above $\tau=1$,
the null point is located just above the temperature minimum ($z\ge400
\;\textrm{km}$), right at the base of the chromosphere.

The magnetic field connectivity around the null point may be regarded
as being a typical quadrupolar configuration in 2D, here an EW
oriented $-;+;-;+$, where the first $+$ polarity is the positive flux
concentration of the EFR, and the second $+$ is the extended western
positive polarity of the active region. The magnetic configuration,
however, is fully 3D. Understanding the sequence of reconnection
events that may have led to the observed activity requires a 3D
analysis.

As for all 3D null points, a fan surface and a singular spine field
line originate from the null. On the one hand, the fan surface forms a
dome which encloses the positive polarity of the EFR (surrounded by a
closed polarity inversion line as shown in Fig.~\ref{fig13}a). On the
other hand, the spine connects the EFR to the distant western positive
polarity of the active region. For relatively low $\alpha$ values, the
spine field line (which corresponds to the first eigenvector of the
null point) is EW oriented and almost parallel to the H$\alpha$ 
surges (compare the dark blue line in Fig.~\ref{fig13}a and the surge in
Fig.~\ref{fig3}).

The null point is a so-called ``improper null'' (i.e., it is
non-axisymmetric around the spine axis, due to the different values of
its two fan-related eigenvalues). In all the extrapolations, the
eigenvalue of the eigenvector oriented in the NS direction is several
times smaller than that one oriented in EW direction, so that magnetic
field lines just above or below the fan surface are predominantly EW
oriented. This static configuration is similar to that described
by \citet{Masson:09}. The structure of the fan is very robust to
changing the lower boundary conditions (the magnetogram) or the
force-free parameter $\alpha$ for a given magnetogram. The former is a
natural consequence of the strength of the overlying EW oriented
bipolar field, which does not change significantly with time; the
latter is due to the well-known insensitivity of small-scale loops to
$\alpha$-variations in the linear force-free field approximation
\citep{Alex:81}.

\subsection{Interpretation of the H$\alpha$ surges and
\ion{Ca}{2} H brightenings}

To understand the origin of the observed surges and chromospheric
brightenings, one first needs to know whether the null point topology
found in the extrapolations existed prior to the event (as, e.g., in
the jet model by \citealt{Pariat:10}) or if it was dynamically formed
while the surges were launched (as, e.g., in the so-called two-step
reconnection model by \citealt{Torok:09} or in the emergence-driven
jet model by \citealt{Moreno:08}). The SST magnetograms clearly show
that a positive flux concentration was present in the dispersed
negative polarity of the active region several tens of minutes before
the appearance of the first surge (see Fig.~\ref{fig5} at 8:58 UT).
The same feature can be observed since at least 7:50 UT in the
magnetograms taken by \emph{Hinode} \citep{Guglielmino:08}. The early
existence of this parasitic polarity and the finding of a null point
above it in all the extrapolations imply that the surges were
triggered in a pre-existing null point topology.

The SST magnetograms shows the emergence of a new small bipole at about 
9:18 UT, right before the first surge. It appears at the NE of the
main positive polarity of the EFR. Until 9:25 UT, the bipole has an
elongated shape with curly flux concentrations that look like those
found in various MHD simulations of the emergence of sub-photospheric
twisted flux ropes. This indicates the local presence of magnetic
shear, and maybe twist, above the photosphere. Because of the shear/twist, 
the potential and LFFF extrapolations fail to connect the two polarities 
of the elongated structure. However, since the bipole is oriented 
in the NE-SW direction, we know that the emerging field 
lines make an angle of several tens of degrees with the
overlaying field lines (see the green and cyan fan field lines 
of the null point in Fig.~\ref{fig13}a).

Fig.~\ref{fig13}b shows an EW oriented red field line that overlays
the EFR and is located beneath the fan surface of the null point.
Shortly after the beginning of the emergence of the twisted bipole,
the lowermost growing sheared field lines could have pushed upwards
this pre-existing fan-related red arcade. As a consequence, the red
arcade would have reconnnected with the long pink field lines located
westward of it, forming new connections: one long field line (green in
Fig.~\ref{fig13}b) and one short arcade (yellow in Fig.~\ref{fig13}b).
In this scenario, H$\alpha$ surges would occur along the green field
line (which is close to, but slightly south from, the spine of the
null point), and bright \ion{Ca}{2} H flare-like loops would form at
the position of the yellow field lines (which are south from the
pre-existing positive flux concentration). This is exactly what is
seen in the observations.

The scenario described above may also explain why several surges are
observed, and why the \ion{Ca}{2} H emission gradually increases with
time.  Indeed, as more and more sheared/twisted field lines emerge,
all the overlying, nearly-potential arcades would reconnect, leaving
room for the sheared field lines themselves to undergo reconnection
processes.  Also, since the time sequence of magnetograms suggests the
rise of a sub-photospheric twisted flux rope, the emerging field lines
must be more and more sheared with time. This means that also the
reconnection must involve field lines that are more and more sheared.
As a consequence, non-smooth vertical shear profiles in the
emerging/reconnecting field lines should naturally lead to non-smooth
reconnection sequences, therefore to distinct surges. In addition, the
reconnections should become more energetic with time since they
involve field lines that contain more and more free magnetic energy,
thus leading to brighter emissions in the \ion{Ca}{2} H line core.

One may argue that this set of (red; pink, green; yellow) field lines
cannot reconnect -- and thus that our interpretation is invalid --
because it is not exactly located at the null point. Two
counter-arguments can be given. First, the exact position of the null
point is not known, as it is sensitive to variations of the $\alpha$
parameter. If the field were non-potential, the null point could 
have been located right on the side of the red arcade. Second, it has been
shown by \cite{Masson:09} that, under some conditions, a (red) field
line initially anchored several Mm away from a null point can still
reconnect at it, and that the (green) reconnected field line can
eventually reach a position away from the spine of the null, as these
lines can ``slip-run'' towards and away from it. The condition found
by \cite{Masson:09} for this to happen is the existence of an
elongated quasi-separatrix layer in which the spine is embedded, which
is associated with the asymmetric property of the null point. As the
null point is ``improper'' in all the extrapolations, the condition
for the occurence of coupled slip-running/null-point reconnection is
satisfied.

\section{Summary}

In this paper we have studied the configuration and evolution of a
small-scale EFR using very high-resolution filtergrams taken at the
SST. We have also analyzed simultaneous observations performed by the
three telescopes aboard \emph{Hinode}. The measurements cover a period
of $\sim$2 hours and a broad wavelength range from the visible to soft
X-rays.

These data have allowed us to observe the flux emergence event in its
entirety, and to study the characteristics of the EFR and its temporal
evolution at different heights in the solar atmosphere.

The photospheric evolution of the emerging flux in the G band is
characterized by the distortion of the granulation pattern and the
appearance of transient dark lanes. The dark lanes are associated 
with horizontal fields and upflows \citep{Guglielmino:08}. This
confirms recent high-resolution observations and numerical
simulations. Moreover, the sequence of \ion{Fe}{1} magnetograms shows
that the field does not emerge as a monolithic bipole, but as a series
of small-scale, mixed-polarity flux patches.

Various phenomena indicating the release of energy and consequent
plasma heating, as well as the formation of jets, occurred at the EFR
site: brightenings in \ion{Ca}{2} H, H$\alpha$, and EUV lines,
intensity enhancements in X-rays, and H$\alpha$ surges.  We have
analyzed the spatial and temporal relationship between all these
features, in order to clarify the evolution of the underlying process.

The brightness enhancements were first seen in the chromospheric
\ion{Ca}{2} H line, which has the lowest formation height. The
beginning of the \ion{Ca}{2} H enhancement occurred at the time the
flux started to increase in the upper photosphere, around 9:10 UT. An
H$\alpha$ surge was observed at 9:23 UT, and the \ion{Ca}{2} H
intensity continued to grow until a localized, intense brightening
appeared at 9:27 UT. The H$\alpha$ intensity began to rise at 9:27 UT,
producing a well-defined brightening at 9:39 UT, so there was a
temporal delay of more than 10 minutes with respect to \ion{Ca}{2} H.
This is consistent with the higher formation height of the H$\alpha$
measurements. However, the H$\alpha$ and \ion{Ca}{2} H brigthness
peaked almost simultaneously, around 9:45 UT. This was also the time
of the maximum X-ray intensity detected with \emph{Hinode} at the
position of the EFR. Two more H$\alpha$ surges appeared during the
emergence event, at around 9:37 and 9:52 UT. The observed time lags
support the interpretation of all these phenomena as being caused by
magnetic reconnection at low altitude in the atmosphere.

The EIS measurements show localized brightenings in the coldest EUV
lines, with small flows that are not significant or extend over much
larger areas. The brightness enhancement decreases with increasing
ion temperature (i.e., formation height) of the lines. Moreover, we 
detect strong non-thermal broadenings except in the hot \ion{Fe}{12} 
195 \AA\/ line. 

We have investigated the relationship between the evolution of the
flux and the onset of jet-like activity. \citet{Yoshimura:03} found
that surges appear during the declining phase of the magnetic flux,
due to cancellation between the emerging flux and the pre-existing
fields of opposite polarity.  On the contrary, in our observations the
surges set up while the flux was still increasing in the upper
photosphere (see Fig.~\ref{fig4}). This suggests that the magnetic
reconnections that supplied the energy of the surges occurred above the
formation height of the \ion{Na}{1} D1 magnetograms, probably in the
low chromosphere.

All these findings point to the emergence event as the likely trigger
of the reconnection process, due to the interactions between two flux
systems: the pre-existing arcades and the EFR field lines. In order to
shed light on this issue, we have performed LFFF extrapolations to
determine the coronal connectivity above the EFR and the possible
reconfiguration of the ambient and emerging field lines.

The scenario suggested by the extrapolations is the following: as the
positive polarity of the EFR appears on the surface with its
conjugate negative polarity, the footpoint separation increases due to
the diverging motion, so the emerging lines bulge upwards. They
therefore push the overlying arcades, which can reconnect to produce 
a set of long loops (reminiscent of the H$\alpha$ surges) and smaller
loops (reminiscent of the observed small brightening regions). A
fan-spine topology forms, and its structure is very stable against
changes in the extrapolation parameters, like the $\alpha$-value or
the boundary conditions.

The 3D null point lies quite low in the atmosphere, $150-400 \;\textrm{km}$ above
the formation height of the \ion{Fe}{1} $630.25\;\textrm{nm}$ line.
Thus, it is located at the base of the chromosphere ($\sim 500
\;\textrm{km}$). The observed H$\alpha$ surges are nearly cospatial
with field lines produced by the reconnection of arcades overlying the
EFR and other active region loops.  Therefore, they could represent
reconnection jets formed by accelerated material propagating along
those lines. The emerging field lines would not participate directly
in the surges, but would indirectly trigger them by pressing the
ambient fields together and forcing their reconnection.

\citet{Torok:09} modeled the emergence of a twisted flux tube into a
potential field arcade, showing that a 3D null point topology is
produced in the corona. This configuration results from a two-step
reconnection process, which transfers the stress of the emerging
magnetic lines (twist or shear) into the coronal arcades by means of a
torsional Alfv\'en wave, with the onset of a jet. In our case, it
seems that the null point had already formed when the SST started to
observe -- most probably as a result of the emergence of the very 
first positive flux patches --, so the situation is closer to that modeled
by, e.g., \citet{Pariat:10}. We thus favor the simple explanation
given above as opposed to a more complex two-step reconection process,
although the former cannot be ruled out completely.

Our analysis combines multi-wavelength observations and detailed
magnetic field extrapolations. This has enabled us to understand the
complex sequence of events associated with the EFR as being the result
of magnetic reconnection triggered by the emergence of new flux into
the atmosphere. Processes like the one considered here are difficult
to interpret without the help of high resolution measurements and
extrapolations. Both techniques have reached a sufficient degree of
maturity, so we believe that this kind of analyses will become standard
practice in the near future.

\acknowledgments

It is a pleasure to thank V. Domingo for his help with the SST
observations, J.A. Bonet for advice concerning the MOMFBD technique,
I. Pagano for useful suggestions in the UV analysis, and all the
scientists of the \emph{Hinode} team for the operation of the
instruments.  This work was started while one of us (S. Guglielmino)
was a Visiting Scientist at the Instituto de Astrof\'isica de
Andaluc\'ia. Financial support by the European Commission through the
SOLAIRE Network (MTRN-CT-2006-035484) is gratefully acknowledged. This
research has been partly funded by the Spanish Ministerio de Ciencia e
Innovaci\'on through projects ESP2006-13030-C06-02, PCI2006-A7-0624,
and Programa de Acceso a Infraestructuras Cient\'ificas. The Swedish 1-m 
Solar Telescope is operated by the Institute for Solar Physics of
the Royal Swedish Academy of Sciences in the Spanish Observatorio del
Roque de los Muchachos of the Instituto de Astrof\'isica de
Canarias. \emph{Hinode} is a Japanese mission developed and launched
by ISAS/JAXA, with NAOJ as domestic partner and NASA and STFC (UK) as
international partners. It is operated by these agencies in
co-operation with ESA and NSC (Norway). Use of NASA's Astrophysical
Data System is gratefully acknowledged.

{\it Facilities:} \facility{SST}, \facility{Hinode}.

\end{document}